\documentclass[twocolumn,pre]{revtex4}
\usepackage{amsmath}
\usepackage{amssymb}
\usepackage{graphicx}
\usepackage{esint}

\begin{document}

\title{Multi-dimensional Biochemical Information Processing of Dynamical
Patterns}

\author{Yoshihiko Hasegawa}

\email{hasegawa@biom.t.u-tokyo.ac.jp}

\date{\today}

\affiliation{Department of Information and Communication Engineering, Graduate
School of Information Science and Technology, The University of Tokyo,
Tokyo 113-8656, Japan}
\begin{abstract}
Cells receive signaling molecules by receptors and relay information
via sensory networks so that they can respond properly depending on
the type of signal. Recent studies have shown that cells can extract
multi-dimensional information from dynamical concentration patterns
of signaling molecules. We herein study how biochemical systems can
process multi-dimensional information embedded in dynamical patterns.
We model the decoding networks by linear response functions, and
optimize the functions with the calculus of variations to maximize
the mutual information between patterns and output. We find that,
when the noise intensity is lower, decoders with different linear
response functions, i.e., distinct decoders, can extract much information.
However, when the noise intensity is higher, distinct decoders do
not provide the maximum amount of information. This indicates that,
when transmitting information by dynamical patterns, embedding information
in multiple patterns is not optimal when the noise intensity is very
large. Furthermore, we explore the biochemical implementations of
these decoders using control theory and demonstrate that these decoders
can be implemented biochemically through the modification of cascade-type
networks, which are prevalent in actual signaling pathways. 

\end{abstract}
\maketitle

\section{Introduction}

Cells receive signals by receptors and subsequently process the obtained
information through biochemical networks so that they can respond
properly. In addition to static information, such as the concentration
or identity of signaling molecules, recent experimental evidence shows
that cells can process dynamical patterns \cite{Behar:2010:TemporalCodes,Purvis:2013:SignalingReview}.
Specifically, it was reported that biochemical networks can filter
dynamical signals in order to counteract noise or for prediction \cite{Kobayashi:2010:PRL,Hinczewski:2014:CellFilter,Becker2015:CellPrediction}.
Because one-dimensional \emph{static} signals can be specified by
a single variable (e.g., the concentration), they provide only one-dimensional
information. On the other hand, one-dimensional \emph{dynamical} signals
require multi-dimensional information to specify their shape, and
hence they are multi-dimensional. The extraction of the dynamical
patterns lets cells learn more about the environment. For multicellular
organisms, dynamical patterns are used for inter-cellular communication.
A biophysical example of inter-cellular information transmission using
dynamic patterns is insulin \cite{Purvis:2013:SignalingReview}. Based
on experiments, it has been reported that multiple messages are embedded
in dynamical patterns and that each specific pattern is selectively
decoded by their downstream molecular networks \cite{Kubota:2012:InsulinAKT,Noguchi:2013:Insulin,Sano:2016:DynPattSig}.
One notable advantage of using dynamical patterns for communications
over static patterns is considered to be the ability to encode more
information into a common molecular species \cite{Selimkhanov:2014:DynSig}.
Although cellular dynamical information processing has attracted much
attention \cite{Tostevin:2009:MI,Kobayashi:2010:PRL,Mora:2010:MLE,Mugler:2010:OscSignal,Kubota:2012:InsulinAKT,Purvis:2012:Insulin,Hansen:2013:DynDec,Noguchi:2013:Insulin,Behar:2013:Dynamics,Hinczewski:2014:CellFilter,Selimkhanov:2014:DynSig,McMahon:2015:MI,Becker2015:CellPrediction,Makadia:2015:DynamicSignal,Sano:2016:DynPattSig},
very little attention has been paid to the multi-dimensional aspects
of the information processing of dynamical patterns \cite{Ronde:2011:Multiplex,Ronde:2014:Multiplex,Selimkhanov:2014:DynSig}. 

\begin{figure}
\includegraphics[width=9cm]{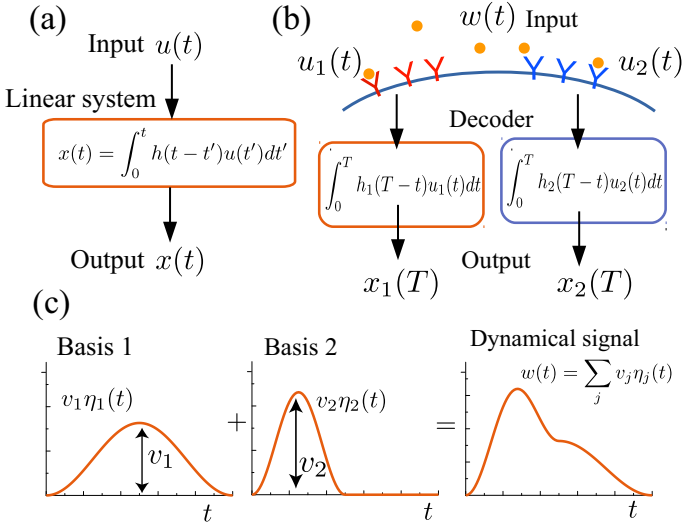}\protect\caption{(a) Relation between input $u(t)$ and output $x(t)$ in a linear
system through linear response function $h(t)$. (b) Decoding of dynamical
patterns by two decoding systems. Input signal $w(t)$ is received
by $N$ types of receptors, and the signal is processed by subsequent
internal molecular decoders ($N=2$ in this figure). The linear response
function of the $i$th decoder is given by $h_{i}(t)$. Each decoder
outputs results by $x_{i}(T)$. (c) Dynamical pattern $w(t)$ as a
sum of basis functions $\eta_{j}(t)$ at intensities $v_{j}$. \label{fig:diagram}}
\end{figure}

Here, we study how biochemical systems can optimally extract multi-dimensional
information from dynamical patterns. Considering the deterministic
limit of decoders (vanishing intrinsic noise limit), we can describe
their response by linear response functions (Fig.~\ref{fig:diagram}(a)).
For dynamical signals with two basis functions and two types of decoders,
we obtain an optimal linear response function through the calculus
of variations in order to maximize mutual information between dynamical
patterns and output. We find that decoders with different linear response
functions (\emph{distinct }decoders) can achieve optimal extraction
of the information from dynamical patterns. However, when the noise
intensity is excessively high, the use of decoders with the same linear
response function (\emph{identical} decoders) can extract more information
than the use of distinct decoders. Using control theory, we also show
that these optimal decoders can be implemented biochemically by a
cascade-type linear signaling network with additional feedforward
and feedback loops, which are prevalent in actual signaling pathways.

\begin{figure}
\includegraphics[width=7cm]{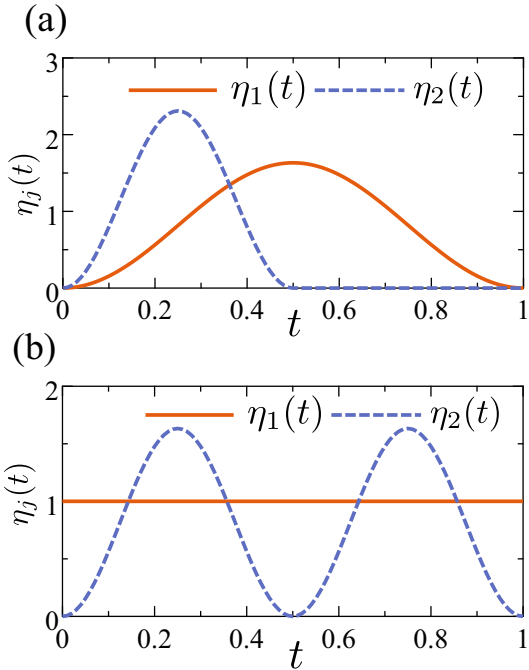}

\protect\caption{Basis sets for the $M=2$ case {[}solid and dashed lines denote $\eta_{1}(t)$
and $\eta_{2}(t)$, respectively{]}. (a) Slow and fast patterns (basis
set A). (b) Constant and oscillation patterns (basis set B). \label{fig:BasisSet}}
\end{figure}

\section{Models}

We consider a biochemical sensory system that reads out extracellular
dynamical patterns by receptors, subsequently processes the signal
via decoding networks, and finally reports the result as the concentration
of output molecular species {[}Fig.~\ref{fig:diagram}(b){]}. We
assume that there exist $N$ decoding systems, each of which consists
of receptors and a subsequent decoding network {[}$N=2$ for Fig.~\ref{fig:diagram}(b){]}.
In order to quantify the amount of transmitted information, we need
to define the probability density on dynamical patterns. As each dynamical
pattern has infinite dimensions, the definition of their probability
density function is not trivial. We model a dynamical pattern $w(t)$
by a sum of basis functions after Ref.~\cite{Gallager:1968:InfoTheory}:
\begin{equation}
w(t)=\sum_{j=1}^{M}v_{j}\eta_{j}(t),\label{eq:wt_def}
\end{equation}
where $M$ is the number of basis functions, $\boldsymbol{\eta}(t)=[\eta_{1}(t),...,\eta_{M}(t)]$
are basis functions, and $\boldsymbol{v}=[v_{1},...,v_{M}]$ are their
coefficients, which are referred to as intensities. Figure~\ref{fig:diagram}(c)
describes the model, where a dynamical pattern $w(t)$ is composed
of two basis functions, $\eta_{1}(t)$ and $\eta_{2}(t)$. The basis
functions need not be orthogonal. However, except for a particular
case of $I^{\mathrm{dup}}$ considered later herein, the basis functions
should be linearly independent. We define probability density $P(\boldsymbol{v})$
on $\boldsymbol{v}$, which are used to define the probability density
of the dynamical patterns. Although Eq.~\eqref{eq:wt_def} is introduced
to incorporate the probability density on dynamical patterns, the
basis functions $\eta_{j}(t)$ and their number $M$ have direct biological
interpretations for some intercellular communication. Cells can decode
multiplexed dynamical patterns \cite{Behar:2010:TemporalCodes,Kubota:2012:InsulinAKT,Purvis:2012:Insulin,Behar:2013:Dynamics,Purvis:2013:SignalingReview,Sano:2016:DynPattSig},
where the patterns can be broadly classified into two basic dynamics,
fast pulsatile and slow transient patterns. Cells can read out amplitude
information embedded in the two patterns. In this example, the number
of basis functions is $M=2$, and $\eta_{1}(t)$ and $\eta_{2}(t)$
correspond to the fast and slow patterns.

We assume that $\eta_{j}(t)$ is in a steady state for $t<0$, where
we define $\eta_{j}(t)=0$ for the steady state concentration (and
hence $w(t)=0$ for $t<0$), and $w(t)$ starts to change at time
$t=0$. Due to stochasticity, accompanied by, e.g., stochastic receptor-ligand
binding, each decoder reads out a degraded pattern $u_{i}(t)$: 
\[
u_{i}(t)=w(t)+\xi_{i}(t),
\]
where $\xi_{i}(t)$ is the input noise of the $i$th type of receptors,
defined by $\left\langle \xi_{i}(t)\right\rangle =0$, and $\left\langle \xi_{i}(t)\xi_{i'}(t^{\prime})\right\rangle =2D_{i}\delta_{ii'}\delta(t-t^{\prime})$,
where $D_{i}$ is the noise intensity. The noise intensity $D_{i}$
depends primarily on the number of $i$th-type receptors and the dissociation
constant of the binding-unbinding reaction \cite{Wang:2007:QuantifyNoise}.
Moreover the dissociation constant has a temperature dependence via
the van't Hoff equation. 

Next, we model the dynamics of the decoders. Let $x_{i}(t)$ be the
output concentration of the $i$th decoder at time $t$, and, for
$t<0$, we define $x_{i}(0)=0$. Note that $w(t)$ and $x_{i}(t)$
are concentrations relative to steady state and so can take negative
values. In order to make analytic calculation possible, we consider
a deterministic limit of decoders \cite{Govern:2012:LinearNetwork,Becker2015:CellPrediction}.
Decoders consist of biochemical reactions subject to intrinsic noise,
the concentration dynamics of which can be described by stochastic
processes. If we consider an infinitely large number of molecules
while keeping the concentration constant, intrinsic noise vanishes
and the stochastic processes reduce to deterministic differential
equations, which is referred to as the deterministic limit. We assume
that decoders output results after a finite time $t=T$ (for simplicity,
we set the same time interval for each decoder), and so $\boldsymbol{x}(T)=[x_{1}(T),...,x_{N}(T)]$
contain information on the dynamical pattern. Suppose that the $i$th
decoder is the single-layer linear decoder (linear birth-death process)
given by 
\begin{equation}
\dot{z}_{i}(t)=-\theta_{i}z_{i}+u_{i}(t),\label{eq:zi_def}
\end{equation}
where $z_{i}(t)$ is the concentration of molecular species in the
$i$th decoder, and $\theta_{i}$ is the degradation rate. In this
decoder, $z_{i}(t)$ directly reports the result, i.e., $x_{i}(T)=z_{i}(T)$.
A similar model was proposed for decoding calcium oscillation \cite{Marhl:2006:CaDecoder}.
Because of the linearity of Eq.~\eqref{eq:zi_def}, the output at
time $t$ is given by a convolution integral: 
\begin{equation}
x_{i}(t)=\int_{0}^{t}h_{i}(t-t^{\prime})u_{i}(t^{\prime})dt^{\prime},\label{eq:xit_conv_def}
\end{equation}
where $h_{i}(t)$ is the linear response function. For this single-layer
and linear case, $h_{i}(t)=e^{-\theta_{i}t}$. Biochemical decoders
are often composed of multiple layers, which can yield complex linear
response functions $h_{i}(t)$ {[}cf.~Eq.~\eqref{eq:LTI_def}{]}
\cite{Govern:2012:LinearNetwork,Hinczewski:2014:CellFilter,Becker2015:CellPrediction}.
For arbitrary linear response functions, the average at $t=T$ is
$\mu_{x_{i}}=\left\langle x_{i}(T)\right\rangle =\int_{0}^{T}h_{i}(T-t^{\prime})w(t^{\prime})dt^{\prime}$,
and the variance is $\sigma_{x_{i}}^{2}=2D_{i}\int_{0}^{T}h_{i}(t^{\prime})^{2}dt^{\prime}$
\cite{Gallager:1968:InfoTheory} (see Appendix~\ref{sec:mean_and_variance}).
Although we used $x(T)$ as the output of the decoders in the present
model, based on Eq.~\eqref{eq:xit_conv_def}, $x(T)$ can also be
regarded as the (weighted) time integration of some intermediate concentration.

Let $x_{i}^{T}=x_{i}(T)$, which is output of the $i$th decoder at
time $t=T$. The amount of information contained in the output $\boldsymbol{x}^{T}=[x_{1}^{T},x_{2}^{T},...,x_{N}^{T}]$
is quantified by the mutual information 
\begin{equation}
I[\boldsymbol{x}^{T};\boldsymbol{v}]=\int d\boldsymbol{x}^{T}\int d\boldsymbol{v}\,P(\boldsymbol{x}^{T}|\boldsymbol{v})P(\boldsymbol{v})\ln\left[\frac{P(\boldsymbol{x}^{T}|\boldsymbol{v})}{P(\boldsymbol{x}^{T})}\right].\label{eq:Ixv_def}
\end{equation}
Here, $P(\boldsymbol{x}^{T}|\boldsymbol{v})$ is the probability density
of $\boldsymbol{x}^{T}$ given $\boldsymbol{v}$, and $P(\boldsymbol{v})$
is the probability density on $\boldsymbol{v}=[v_{1},..,v_{M}]$.
Equation~\eqref{eq:Ixv_def} is the quantity defined between $\boldsymbol{x}$
at time $t=T$ and $\boldsymbol{v}$. We assume independent probability
densities for $v_{j}$, $P(\boldsymbol{v})=\prod_{j=1}^{M}P(v_{j})$,
where $P(v_{j})$ is the Gaussian distribution with mean $0$ and
variance $\sigma_{v_{j}}^{2}$. Although we assumed independence for
$v_{j}$, we can eliminate this assumption when $\boldsymbol{v}$
is distributed according to a multivariate Gaussian distribution.
If $\boldsymbol{v}$ has a multivariate Gaussian distribution, we
can apply a linear transform to redefine basis functions $\boldsymbol{\eta}(t)$
so that elements of $\boldsymbol{v}$ become uncorrelated with each
other (see Appendix~\ref{sec:independence_of_v}). Since uncorrelated
Gaussian random variables are independent, we can always make the
independence assumption for $\boldsymbol{v}$. 

We wish to find optimal decoders which maximally extract information
from dynamical patterns. Instead of exploring all possible candidate
structures, we optimize a set of linear response functions $\boldsymbol{h}(t)=[h_{1}(t),...,h_{N}(t)]$
with the calculus of variations. Thus we obtain a desirable biochemical
system through an optimization problem with an identifiable objective
function \cite{Hasegawa:2013:OptimalPRC,Hasegawa:2014:PRL,Hasegawa:2016:DynSignal}. 

Taking into account biological situations, we consider the following
three optimization problems (italicized words in parentheses are abbreviations):
(i) maximization of $I[\boldsymbol{x}^{T};\boldsymbol{v}]$ (\emph{full
decoder}), (ii) maximization of $I[\boldsymbol{x}^{T};\boldsymbol{v}]$
with $P(\boldsymbol{x}^{T})=\prod_{i}P(x_{i}^{T})$ (\emph{decorrelating
decoder}), and (iii) maximization of $I[\boldsymbol{x}^{T};\boldsymbol{v}]$
with single-layer linear decoders (\emph{SLL decoder}). For (i), decoders
obtained by full maximization provide an upper bound on the mutual
information between dynamical patterns and output. When cells want
to extract as much information as possible, this maximization is suitable.
For (ii), $P(\boldsymbol{x}^{T})=\prod_{i}P(x_{i}^{T})$ can be easily
incorporated into the maximization if $N=M$, which is assumed here.
As the input noises $\boldsymbol{\xi}(t)=[\xi_{1}(t),...,\xi_{N}(t)]$
affect each receptor independently (Fig.~\ref{fig:diagram}(b)),
we have $P(\boldsymbol{x}^{T}|\boldsymbol{v})=\prod_{i=1}^{N}P(x_{i}^{T}|\boldsymbol{v})$.
Combining these relations, we arrive at
\begin{align*}
P(\boldsymbol{x}^{T}) & =\int d\boldsymbol{v}\,P(\boldsymbol{x}^{T}|\boldsymbol{v})P(\boldsymbol{v}),\\
 & =\int d\boldsymbol{v}\,\prod_{i=1}^{N}P(x_{i}^{T}|\boldsymbol{v})\prod_{j=1}^{N(=M)}P(v_{j}).
\end{align*}
 If each $P(x_{i}^{T}|\boldsymbol{v})$ disjointly depends on only
one $v_{i}$ {[}i.e., $P(x_{i}^{T}|\boldsymbol{v})=P(x_{i}^{T}|v_{i})${]},
we can show that $P(\boldsymbol{x}^{T})=\prod_{i}P(x_{i}^{T})$. This
is similar to a decorrelator in digital communication, which decorrelates
multiplexed signals (see Appendix~\ref{sec:optimal_function}). For
this case, $v_{i}$ can be obtained by measuring only one $x_{i}(T)$,
i.e., $I[\boldsymbol{x}^{T};\boldsymbol{v}]=\sum_{i=1}^{N}I[x_{i}^{T};v_{i}]$.
For (iii), we fix the linear response function to $h_{i}(t)=e^{-\theta_{i}t}$,
which corresponds to the abovementioned single-layer linear (SLL)
decoder. We optimize all $\theta_{i}$ numerically with simulated
annealing to maximize $I[\boldsymbol{x}^{T};\boldsymbol{v}]$.

For arbitrary $N$ and $M$ (both $N=M$ and $N\ne M$ are allowed),
we obtain the optimal linear response functions as follows (see Appendix~\ref{sec:optimal_function}):
\begin{equation}
h_{i}(t)=-\frac{1}{4\Lambda_{i}D_{i}}\sum_{j=1}^{M}\lambda_{ij}\eta_{j}(T-t),\label{eq:opt_h}
\end{equation}
where $\lambda_{ij}$ and $\Lambda_{i}$ are Lagrange multipliers
(real values), and these values depend on the type of decoders (full
or decorrelating). When observing the dynamical pattern composed of
a single basis function ($M=1$) with a single decoder ($N=1$), the
optimal linear response function is $h_{1}(t)\propto\eta_{1}(T-t)$,
which is known as the matched filter. From Eq.~\eqref{eq:opt_h},
the optimal linear response function that maximizes the mutual information
is given by the summation of matched filters. Although the matched
filter is known to be optimal for $M=N=1$, the optimality of Eq.~\eqref{eq:opt_h}
for maximization of the mutual information is not trivial.

\section{Results}

\subsection{Mutual information}

We construct concrete optimal linear response functions for a system
with $N=M=2$. In actual inter-cellular communication, as far as
known, the degree of multiplexing is very small. Moreover, obtaining
optimal linear response functions becomes more difficult as $N$ or
$M$ increases. Therefore, we select $N=M=2$ as the minimal model
for the multi-dimensional information processing. For the basis functions
$\eta_{j}(t)$, we consider the two basis sets shown in Figs.~\ref{fig:BasisSet}(a)
and (b). The two basis sets A and B are defined by 
\begin{equation}
\mathrm{Set\,A}\begin{cases}
\eta_{1}(t) & =\sqrt{\frac{2}{3}}(1-\cos(2\pi t)),\\
\eta_{2}(t) & =\frac{2}{\sqrt{3}}\Theta\left(\frac{1}{2}-t\right)(1-\cos(4\pi t)),
\end{cases}\label{eq:setA1_def}
\end{equation}
and
\begin{equation}
\mathrm{Set\,B}\begin{cases}
\eta_{1}(t) & =1,\\
\eta_{2}(t) & =\sqrt{\frac{2}{3}}(1-\cos(4\pi t)).
\end{cases}\label{eq:setB1_def}
\end{equation}
Basis set A comprises slow and fast patterns (Fig.~\ref{fig:BasisSet}(a)),
where $\Theta(t)$ is a step function; and basis set B comprises constant
and oscillation patterns (Fig.~\ref{fig:BasisSet}(b)). All of the
basis functions are normalized so that $\psi_{11}=1$ and $\psi_{22}=1$,
where $\psi_{jj'}$ is the correlation matrix defined by $\psi_{jj'}=\int_{0}^{T}\eta_{j}(t)\eta_{j'}(t)dt$.
Regarding $T$, it is reasonable to choose $T$ as the largest duration
time among $\eta_{j}(t)$. If $T$ is shorter than the largest time,
decoders cannot use all of the information contained in $w(t)$. Conversely,
even if $T$ is longer than the largest time, decoders cannot extract
more information from $w(t)$. Therefore, we use $T=1$ for both of
the basis sets. 

\begin{figure}
\includegraphics[width=9cm]{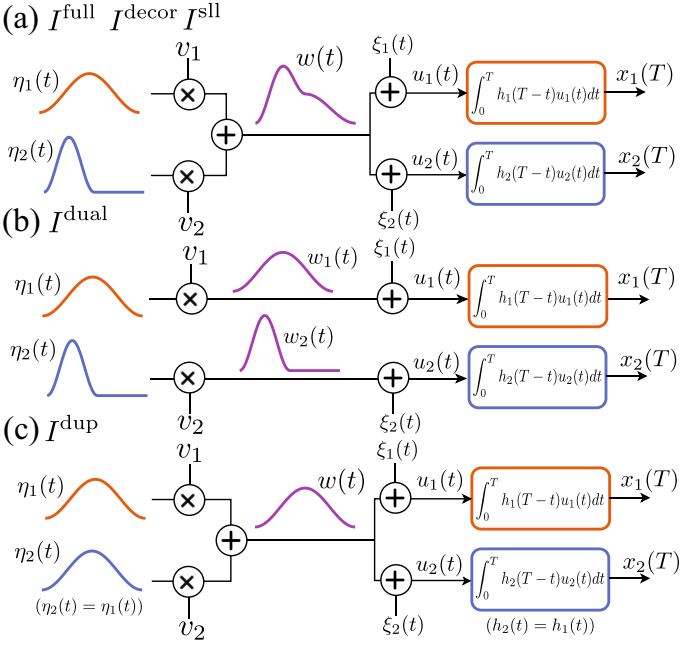}

\protect\caption{Information transmission model assumed in different mutual information.
(a) Information is embedded in two distinct basis functions, transmitted
through one common channel, and decoded by two decoders. (b) Information
is embedded in two distinct basis functions, transmitted through two
designated channels, and decoded by two decoders. (c) Information
is embedded in two identical basis functions, transmitted through
one channel, and decoded by two identical decoders. In (a)--(c), $\oplus$
and $\otimes$ denote addition and multiplication operations, respectively.
\label{fig:information_model}}
\end{figure}

Let $I^{\mathrm{full}}$, $I^{\mathrm{decor}}$, and $I^{\mathrm{sll}}$
be the mutual information $I[\boldsymbol{x}^{T};\boldsymbol{v}]$
of the full, decorrelating, and SLL decoders, respectively. $I^{\mathrm{full}}$,
$I^{\mathrm{decor}}$, and $I^{\mathrm{sll}}$ are obtained by optimizing
the linear response functions. As explained above, the mutual information
quantities assume a model in which information is embedded in two
linearly independent basis functions, is transmitted through a common
channel, and is decoded by two decoders (Fig.~\ref{fig:information_model}(a)).
We cannot obtain closed-form solutions for $I^{\mathrm{full}}$ and
$I^{\mathrm{sll}}$ for arbitrary noise intensity, so we calculate
the solutions numerically (see Appendix~\ref{sec:optimal_function}).
For sufficiently small $D_{1}$ and $D_{2}$, $I^{\mathrm{full}}$
is approximated by 
\begin{equation}
I^{\mathrm{full}}\simeq\frac{1}{2}\ln\left(\frac{\sigma_{v_{1}}^{2}\sigma_{v_{2}}^{2}\left(\psi_{11}\psi_{22}-\psi_{12}^{2}\right)}{4D_{1}D_{2}}\right).\label{eq:I_full_def}
\end{equation}
When $\eta_{1}(t)$ and $\eta_{2}(t)$ are linearly independent, we
have $\psi_{11}\psi_{22}>\psi_{12}^{2}$. $I^{\mathrm{decor}}$ can
be calculated in closed form for arbitrary noise intensity (see Appendix~\ref{sec:optimal_function}).
For sufficiently small $D_{1}$ and $D_{2}$, $I^{\mathrm{decor}}$
is approximated by 
\begin{equation}
I^{\mathrm{decor}}\simeq\frac{1}{2}\ln\left(\frac{\sigma_{v_{1}}^{2}\sigma_{v_{2}}^{2}\left(\psi_{11}\psi_{22}-\psi_{12}^{2}\right)^{2}}{4D_{1}D_{2}\psi_{11}\psi_{22}}\right).\label{eq:I_decor_def}
\end{equation}
For comparison, we consider the mutual information $I^{\mathrm{dual}}$
which corresponds to a model where information embedded in two basis
functions is transmitted through two designated channels and is decoded
by two designated decoders (Fig.~\ref{fig:information_model}(b)).
Applying the calculus of variations, $I^{\mathrm{dual}}$ is represented
by
\begin{alignat}{1}
I^{\mathrm{dual}} & =I[x_{1}^{T};v_{1}]+I[x_{2}^{T};v_{2}],\nonumber \\
 & =\frac{1}{2}\ln\left(1+\frac{\sigma_{v_{1}}^{2}\psi_{11}}{2D_{1}}\right)+\frac{1}{2}\ln\left(1+\frac{\sigma_{v_{2}}^{2}\psi_{22}}{2D_{2}}\right).\label{eq:I_dual_def}
\end{alignat}
Note that $I^{\mathrm{dual}}$ does not have biological relevance
but is introduced merely as a theoretical reference point. From Eqs.~\eqref{eq:I_full_def}
and \eqref{eq:I_dual_def}, when $D_{1}$ and $D_{2}$ are sufficiently
small, the following relation holds: 
\begin{equation}
I^{\mathrm{full}}\le I^{\mathrm{dual}},\label{eq:Ifull_Idual_ineq}
\end{equation}
where it holds with equality when the correlation between the two
basis functions is zero (i.e., $\psi_{12}=0$). 

\begin{figure*}
\includegraphics[width=18cm]{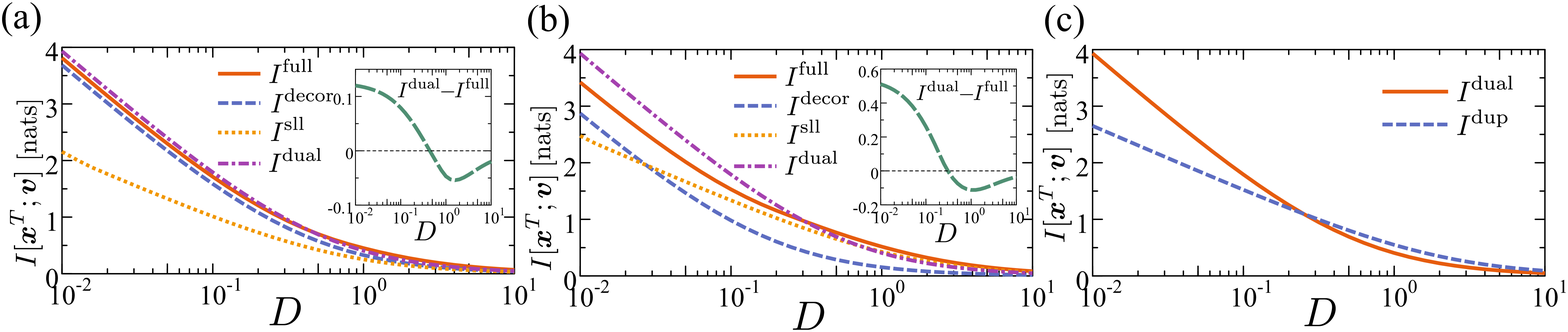}

\protect\caption{(a) and (b) Mutual information as a function of noise intensity $D$
by basis set; (a) basis set A {[}Fig.~\ref{fig:BasisSet}(a){]} and
(b) basis set B {[}Fig.~\ref{fig:BasisSet}(b){]}. Solid, dashed,
dotted, and dot-dashed lines denote $I^{\mathrm{full}}$, $I^{\mathrm{decor}}$,
$I^{\mathrm{sll}}$, and $I^{\mathrm{dual}}$, respectively. In (a)
and (b), the insets highlight $I^{\mathrm{dual}}-I^{\mathrm{full}}$
as a function of $D$. (c) Mutual information quantities $I^{\mathrm{dual}}$
(solid line) and $I^{\mathrm{dup}}$ (dashed line) as functions of
$D$. Note that $I^{\mathrm{dual}}$ and $I^{\mathrm{dup}}$ do not
depend on the basis set. In all panels, parameters are $T=1$ and
$\sigma_{v_{1}}^{2}=\sigma_{v_{2}}^{2}=1$. \label{fig:MI}}
\end{figure*}

Figures~\ref{fig:MI}(a) and (b) show $I^{\mathrm{full}}$, $I^{\mathrm{decor}}$,
$I^{\mathrm{sll}}$, and $I^{\mathrm{dual}}$ (solid, dashed, dotted,
and dot-dashed lines, respectively) as functions of the noise intensity
$D\ (=D_{1}=D_{2})$ for basis sets A and B, respectively. Parameter
details are shown in the caption of Fig.~\ref{fig:MI}. In Figs.~\ref{fig:MI}(a)
and (b), we see that $I^{\mathrm{full}}$ and $I^{\mathrm{decor}}$
yield higher values than $I^{\mathrm{sll}}$ for a lower noise intensity
$D$, especially in Fig.~\ref{fig:MI}(a), which indicates that optimal
linear response functions extract information more efficiently than
SLL decoders. The insets in Figs.~\ref{fig:MI}(a) and (b) highlight
$I^{\mathrm{dual}}-I^{\mathrm{full}}$ as a function of $D$. Interestingly,
for large noise intensity $D$, the two mutual information quantities
$I^{\mathrm{full}}$ and $I^{\mathrm{dual}}$ obey $I^{\mathrm{full}}>I^{\mathrm{dual}}$,
which is the opposite relation to Eq.~\eqref{eq:Ifull_Idual_ineq}
(Eq.~\eqref{eq:Ifull_Idual_ineq} is satisfied for sufficiently \emph{small}
$D$). This relation is nontrivial because the use of designated channels,
which corresponds to $I^{\mathrm{dual}}$, is expected to provide
higher information transmission, as shown by Eq.~\eqref{eq:Ifull_Idual_ineq}.

\begin{figure*}
\includegraphics[width=18cm]{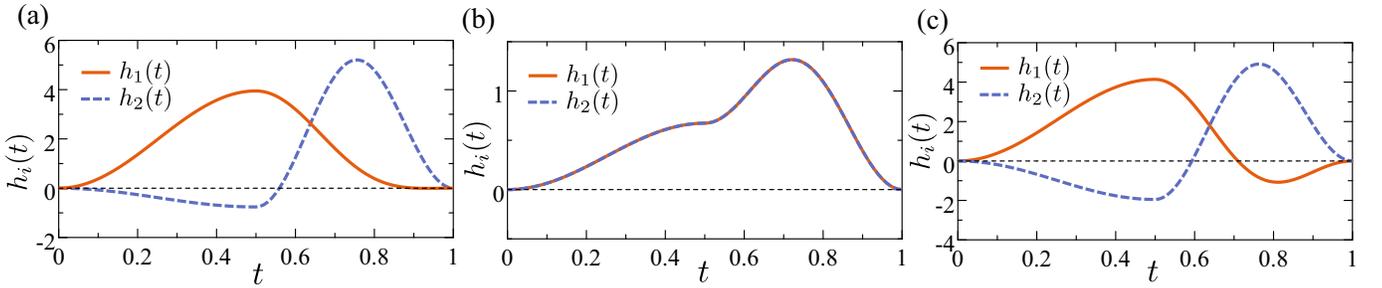}

\protect\caption{(a) and (b) Optimal linear response function $h_{i}(t)$ of the full
decoder with basis set A for different noise intensities; (a) $D=0.1$
and (b) $D=1.0$, where all other parameters are the same. (c) Optimal
linear response function $h_{i}(t)$ of the decorrelating decoder
with $D=0.1$. In all panels, the solid and dashed lines denote $h_{1}(t)$
and $h_{2}(t)$, respectively. For all $h_{i}(t)$ shown, the functionsthat
are horizontally symmetric with respect to $h_{i}=0$ are also optimal
solutions. We set $\sigma_{x_{1}}^{2}=\sigma_{x_{2}}^{2}=1$, and
these parameters affect only the magnitude of the functions. The other
parameters are identical to those in Fig.~\ref{fig:MI}. \label{fig:Linear_MI}}
\end{figure*}

In order to investigate the cause of this opposite relation between
$I^{\mathrm{full}}$ and $I^{\mathrm{dual}}$ with respect to $D$,
we examine the optimal response functions, which are shown in Fig.~\ref{fig:Linear_MI}.
Figures~\ref{fig:Linear_MI}(a) and (b) show linear response functions
$h_{1}(t)$ (solid line) and $h_{2}(t)$ (dashed line) for the full
decoder with basis set A for different noise intensities ($D=0.1$
and $1.0$, respectively) while keeping the other parameters unchanged
(details are shown in the caption of Fig.~\ref{fig:Linear_MI}).
Note that for $D<0.1$, the shapes of the optimal linear response
functions are similar to that of $D=0.1$, and for $D>1.0$, the shapes
are similar to that of $D=1.0$. In Fig.~\ref{fig:Linear_MI}(c),
we also show the optimal linear response functions $h_{1}(t)$ (solid
line) and $h_{2}(t)$ (dashed line) for the decorrelating decoder.
In this case, there is no major difference when noise intensity $D$
is varied. We can see that for $D=0.1$ (Fig.~\ref{fig:Linear_MI}(a)),
the linear response function of the full decoder is similar to that
of the decorrelating decoder of Fig.~\ref{fig:Linear_MI}(c), indicating
that the decorrelation can provide near-optimal efficiency for the
weak-noise case. In Fig.~\ref{fig:Linear_MI}(a), $h_{i}(t)$ indicated
by solid and dashed lines mainly decode information embedded in slow
and fast patterns, respectively. When we increase $D$ in the fully
optimal case, the two linear response functions coalesce to a single
function (the critical points are $D\simeq0.83$ for basis set A and
$D\simeq0.20$ for basis set B). This result indicates that, when
the noise intensity is very strong, decoding information with two
distinct decoders is inefficient but decoding with identical decoders
is relatively efficient. Therefore, in the region in which $I^{\mathrm{full}}$
and $I^{\mathrm{dual}}$ obey $I^{\mathrm{dual}}<I^{\mathrm{full}}$,
a qualitative change in the linear response functions occurred. 

In order to explain this change in great detail, we introduce another
mutual information quantity $I^{\mathrm{dup}}$, which assumes a model
similar to that shown in Fig.~\ref{fig:information_model}(a) but
uses the same function for the two basis functions ($\eta_{1}(t)=\eta_{2}(t)$)
and the same linear response function for the two decoders ($h_{1}(t)=h_{2}(t)$),
as shown in Fig.~\ref{fig:information_model}(c). We also set $D_{1}=D_{2}=D$
and $\sigma_{v_{1}}^{2}=\sigma_{v_{2}}^{2}=\sigma_{v}^{2}$. Optimizing
the linear response functions, $I^{\mathrm{dup}}$ is given by (see
Appendix~\ref{sec:optimal_function}) 
\begin{align}
I^{\mathrm{dup}} & =\frac{1}{2}\ln\left(1+\frac{2\sigma_{v}^{2}\psi}{D}\right),\label{eq:I_dup_def_1}
\end{align}
where $\psi=\psi_{11}=\psi_{22}$. Figure~\ref{fig:MI}(c) shows
$I^{\mathrm{dual}}$ and $I^{\mathrm{dup}}$ as functions of $D$,
and we observe that $I^{\mathrm{dual}}>I^{\mathrm{dup}}$ for weaker
noise intensity, while $I^{\mathrm{dup}}>I^{\mathrm{dual}}$ for larger
noise intensity. This indicates that when the noise intensity is excessively
large, multi-dimensional information transmission becomes inefficient.
Transmitting information by embedding information into two identical
basis functions and decoding using two identical decoders becomes
more efficient.

\begin{figure*}
\includegraphics[width=18cm]{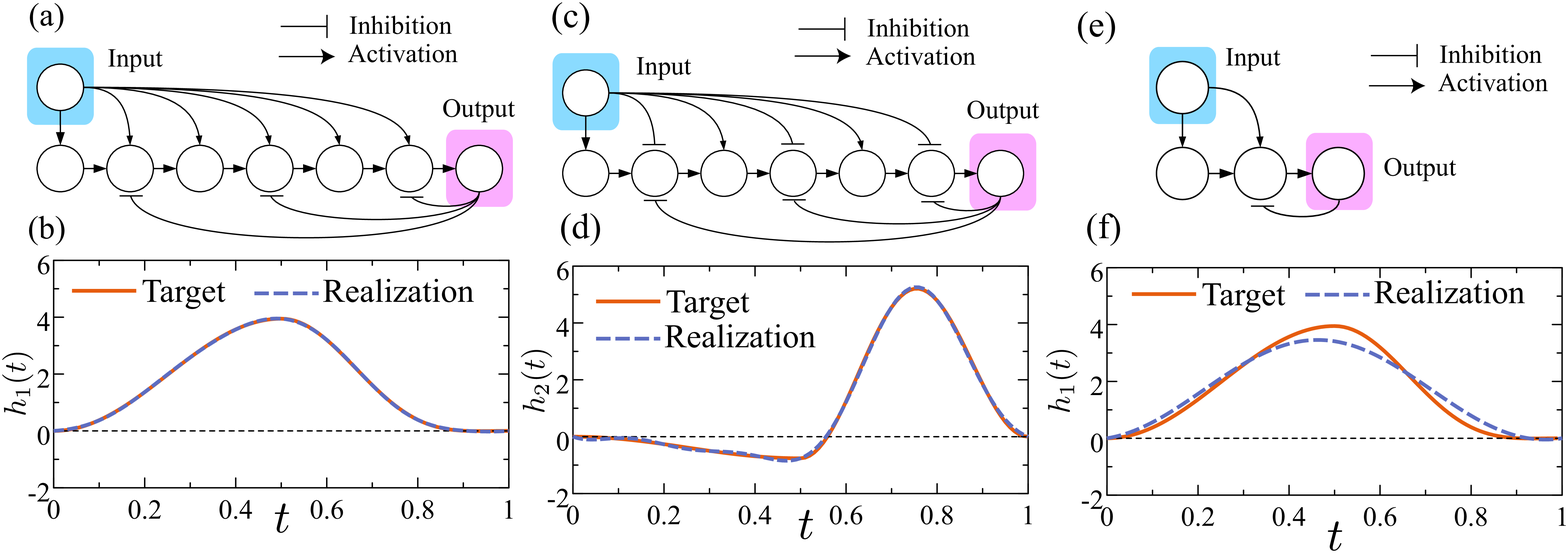}

\protect\caption{(a) Molecular realization of $h_{1}(t)$ in Fig.~\ref{fig:Linear_MI}(a),
and (b) its linear response function. In (a), arrows and bar-headed
arrows denote activation and inhibition, respectively. In (b), the
dashed line is the linear response function of the realization network
and the solid line is the target optimal linear response function
shown in Fig.~\ref{fig:Linear_MI}(a). (c) Molecular realization
of $h_{2}(t)$ in Fig.~\ref{fig:Linear_MI}(a) and (d) its linear
response function. (e) Reduced molecular realization for the full
network shown in Fig.~\ref{fig:network}(a) and (f) its linear response
function. In (c)--(f), the meanings of the arrows and lines are the
same as in (a) and (b). \label{fig:network}}
\end{figure*}

\subsection{Biochemical implementation}

We next explore a biochemical implementation of the optimal decoders.
We attempt to implement a decoding network corresponding to $h_{i}(t)$
with $K_{i}$ molecular species ($K_{i}$ is determined by the degree
of the transfer function; see below). Linearizing around the steady
state, we describe their dynamics by the following linear model: 
\begin{equation}
\dot{\boldsymbol{z}}_{i}(t)=\boldsymbol{A}_{i}\boldsymbol{z}_{i}(t)+\boldsymbol{b}_{i}u_{i}(t).\label{eq:LTI_def}
\end{equation}
where $\boldsymbol{z}_{i}(t)=[z_{i1}(t),....,z_{iK_{i}}(t)]^{\top}$,
$z_{ik}(t)$ is the relative concentration of the $k$th molecular
species in the $i$th decoder, $\boldsymbol{A}_{i}$ is a $K_{i}\times K_{i}$
matrix, and $\boldsymbol{b}_{i}$ is a $K_{i}$-dimensional column
vector. The output of Eq.~\eqref{eq:LTI_def} is $z_{iK_{i}}(t)$
and hence $x_{i}(T)=z_{iK_{i}}(T)$ (the last molecular species reports
the result). Independent of the type of maximization (the full or
decorrelating decoders), from Eq.~\eqref{eq:opt_h}, Laplace transform
yields 
\begin{equation}
\tilde{h}_{i}(s)=-\frac{1}{4\Lambda_{i}D_{i}}\sum_{j=1}^{M}\lambda_{ij}\tilde{\eta}_{j}(s),\label{eq:htilde_def}
\end{equation}
where $\tilde{h}_{i}(s)=\mathcal{L}[h_{i}(t)]$ (the transfer function)
and $\tilde{\eta}_{j}(s)=\mathcal{L}[\tilde{\eta}_{j}(T-t)]$ with
$\mathcal{L}$ being the Laplace transform. We want to identify $\boldsymbol{A}_{i}$
and $\boldsymbol{b}_{i}$ which yield the desired transfer functions
$\tilde{h}_{i}(s)$. This problem is known as the realization problem
in control theory \cite{Williams:2007:LSS}. Let the transfer function
be a rational polynomial function of the form 
\begin{equation}
\tilde{h}_{i}(s)=\frac{\sum_{k=1}^{K_{i}}\beta_{ik}s^{K_{i}-k}}{s^{K_{i}}+\sum_{k=1}^{K_{i}}\alpha_{ik}s^{K_{i}-k}},\label{eq:his_def}
\end{equation}
where $\beta_{ik}$ and $\alpha_{ik}$ are real values, and the degree
of the denominator is larger than that of the nominator (this condition
is called \emph{strictly proper}). From control theory, one possible
realization of this transfer function is (see Appendix~\ref{sec:network_realization})
\begin{equation}
\boldsymbol{A}_{i}=\left[\begin{array}{ccccc}
0 & 0 & 0 & \cdots & -\alpha_{iK_{i}}\\
1 & 0 & 0 & \cdots & -\alpha_{i,K_{i}-1}\\
0 & 1 & 0 & \cdots & -\alpha_{i,K_{i}-2}\\
\vdots & \vdots & \vdots & \ddots & \vdots\\
0 & 0 & 0 & 1 & -\alpha_{i1}
\end{array}\right],\boldsymbol{b}_{i}=\left[\begin{array}{c}
\beta_{iK_{i}}\\
\beta_{i,K_{i}-1}\\
\beta_{i,K_{i}-2}\\
\vdots\\
\beta_{i1}
\end{array}\right].\label{eq:OCF}
\end{equation}
Off-diagonal ones in Eq.~\eqref{eq:OCF} imply that $z_{ik}$ depends
on $z_{i,k-1}$ ($k=2,3,...,K_{i}$), which corresponds to a cascade
topology. When the transfer function is strictly proper, its corresponding
linear systems can be implemented by a cascade network with additional
feedback and feedforward loops. As is well known, the cascade topology
is prevalent in actual signaling networks and additional feedback
and feedforward loops exists in the networks, implying that it is
possible to implement optimal decoders biochemically. 

As an example, we construct biochemical implementations for the decoders
of the basis set A with $D=0.1$ (Fig.~\ref{fig:Linear_MI}(a)).
We show the biochemical networks in Figs.~\ref{fig:network}(a) and
(c), which are realizations of $h_{1}(t)$ and $h_{2}(t)$ in Fig.~\ref{fig:Linear_MI}(a),
respectively. Figures~\ref{fig:network}(b) and (d) show linear response
functions of the networks in Figs.~\ref{fig:network}(a) and (c),
respectively. The realization networks are created by applying the
Fourier series expansion to $\eta_{j}(t)$ and calculating their Laplace
transform (see Appendix~\ref{sec:network_realization}). In Figs.~\ref{fig:network}(a)
and (c), when matrix elements of $\boldsymbol{A}_{i}$ and $\boldsymbol{b}_{i}$
are positive or negative, we display their relation by activation
(arrow) or inhibition (bar-headed arrow), respectively. In Figs.~\ref{fig:network}(b)
and (d), we can see that the linear response functions of the molecular
networks (dashed line) are indistinguishable from the target optimal
linear response function $h_{i}(t)$ (solid line). This indicates
that the biochemical networks which maximally exploit information
from dynamical patterns can be implemented. The network in Fig.~\ref{fig:network}(c)
decodes the fast pattern, while that in Fig.~\ref{fig:network}(a)
decodes the slow pattern. The main difference between these two networks
is that the latter has an incoherent feed-forward loop (iFFL) \cite{Mangan:2003:FFL,Alon:2007:SystBiolBook},
while the former does not. Reference~\cite{Noguchi:2013:Insulin}
indicated that, when decoding temporal insulin patterns, a decoding
network having an iFFL is responsive against a fast pulsatile pattern,
while it does not respond to a slow ramp pattern. Because transfer
functions $\tilde{h}_{1}(s)$ and $\tilde{h}_{2}(s)$ are sums of
$\tilde{\eta}_{1}(s)$ and $\tilde{\eta}_{2}(s)$ with different weighting,
they are rational polynomial functions with the same denominator unless
they can be reduced. If $\tilde{h}_{1}(s)$ and $\tilde{h}_{2}(s)$
have the same denominator, $\boldsymbol{A}_{1}$ and $\boldsymbol{A}_{2}$
of these realizations become identical (cf. Eqs.~\eqref{eq:his_def}
and \eqref{eq:OCF}) and this is a reason why the realization networks
of Figs.~\ref{fig:network}(a) and (c) have the same feedback structure
from the output. Both of the implementations have 7 nodes (i.e., $K_{1}=K_{2}=7$).
However, we note that the molecular networks can be minimized without
losing much of their response. Specifically, we construct a reduced
realization network for the full network shown in Fig.~\ref{fig:network}(a).
Figures~\ref{fig:network}(e) and (f) show the reduced network, which
consists of 3 nodes ($K_{1}=3$), and its corresponding linear response
function, respectively. The meanings of the arrows in Fig.~\ref{fig:network}(e)
and the lines in Fig.~\ref{fig:network}(f) are identical to those
in Figs.~\ref{fig:network}(a) and (b), respectively. From Fig.~\ref{fig:network}(f),
we see that the response of the reduced realization network (dashed
line) is similar to that of the optimal realization network (solid
line). Although the number of nodes in the reduced network (Fig.~\ref{fig:network}(e))
is smaller than in the corresponding full network (Fig.~\ref{fig:network}(a)),
the basic structures are similar: there are positive feedforward loops
from the input and negative feedback loops from the output, and there
is no iFFL (see Appendix~\ref{sec:network_realization}). We constructed
this reduced network heuristically based on the balanced truncation
in control theory. It is worthwhile to develop a systematic reduction
procedure, which would lead to feasible biochemical implementations.

\section{Concluding Remarks}

In this manuscript, we considered the optimal decoding of dynamical
patterns through maximization of mutual information between input
and output. We found that when the noise intensity is relatively low,
the distinct decoders can extract much information, as expected. On
the other hand, when the noise intensity is very high, distinct decoders
cannot achieve the optimal extraction of the information, while identical
decoders can. Although multiplexing is naturally considered to confer
higher information transmission, our results show that this is not
necessarily true for the case in which receptors are subject to strong
noise. Still, we note that when decoding information with the identical
decoders, it is impossible to demultiplex dynamical signals. Therefore,
the decoders can determine the intensity of $v_{1}$ or $v_{2}$ but
cannot identify whether the intensity corresponds to $v_{1}$ or $v_{2}$.
As indicated by several experiments, cells use multiplexed dynamical
patterns to transmit information. If the primary goal of cellular
sensory networks is transmitting as much information as possible,
our results can provide insight into a possible range of the noise
intensity. Furthermore, we investigated the possibility of biochemical
implementation of the optimal decoders and found that such optimal
decoders can be implemented by a modification of the cascade network. 

Recently, extensive research has been conducted in order to construct
relations between thermodynamic cost and mutual information \cite{Parrondo:2015:InfoThermo},
especially in biological contexts \cite{Sartori:2014:Adaptation,Barato:2014:CellInfo}.
In particular, Ref.~\cite{Ouldridge:2017:ThermoCost} studied the
thermodynamic cost of the mutual information between receptors and
readouts using a Markov process. Our model considers the deterministic
limit and hence it ignores the intrinsic thermal noise. When we incorporate
the effect of intrinsic noise, the mutual information between patterns
and output should be bounded above by some thermodynamic cost. Exploration
of this topic is left for future research.

\section*{Acknowledgments}

This work was supported by KAKENHI Grant No.~16K00325 from the Ministry
of Education, Culture, Sports, Science and Technology.

\appendix

\section{Mean and variance of output\label{sec:mean_and_variance}}

We calculate the mean and the variance of output of the $i$th decoder
as follows. As described in the main text, we can express the output
of the $i$th decoder by Eq.~\eqref{eq:xit_conv_def}. The mean at
time $t=T$ is
\begin{align}
\mu_{x_{i}} & =\left\langle x_{i}(T)\right\rangle ,\nonumber \\
 & =\int_{0}^{T}h_{i}(T-t^{\prime})\left\langle w(t^{\prime})+\xi_{i}(t^{\prime})\right\rangle dt^{\prime},\nonumber \\
 & =\int_{0}^{T}h_{i}(T-t^{\prime})w(t^{\prime})dt^{\prime},\nonumber \\
 & =\sum_{j=1}^{M}v_{j}q_{ij},\label{eq:conv_mean}
\end{align}
where we define 
\begin{equation}
q_{ij}=\int_{0}^{T}h_{i}(T-t^{\prime})\eta_{j}(t^{\prime})dt^{\prime}.\label{eq:qij_def}
\end{equation}
Similarly, the variance at time $t=T$ is given by 
\begin{align}
\sigma_{x_{i}}^{2} & =\left\langle x_{i}(T)^{2}\right\rangle -\left\langle x_{i}(T)\right\rangle ^{2},\nonumber \\
 & =\int_{0}^{T}dt^{\prime}\int_{0}^{T}dt^{\prime\prime}\,h_{i}(T-t^{\prime})h_{i}(T-t^{\prime\prime})\left\langle \xi_{i}(t^{\prime})\xi_{i}(t^{\prime\prime})\right\rangle ,\nonumber \\
 & =2D_{i}\int_{0}^{T}h_{i}(t^{\prime})^{2}dt^{\prime}.\label{eq:conv_var}
\end{align}

\section{Independence of $\boldsymbol{v}$\label{sec:independence_of_v}}

We can make elements in $\boldsymbol{v}=[v_{1},...,v_{M}]$ independent
of each other through a change of basis functions $\boldsymbol{\eta}(t)=[\eta_{1}(t),...,\eta_{M}(t)]$.
We define a covariance matrix $\boldsymbol{\mathcal{C}}=\left\langle \boldsymbol{v}^{\top}\boldsymbol{v}\right\rangle $,
the elements of which are 
\[
\mathcal{C}_{ij}=\left\langle v_{i}v_{j}\right\rangle =\int v_{i}v_{j}P(v_{i},v_{j})dv_{i}dv_{j},
\]
where we assumed $\left\langle v_{j}\right\rangle =0$. Because the
covariance matrix $\boldsymbol{\mathcal{C}}$ is real symmetric, it
can be diagonalized by an orthogonal matrix $\boldsymbol{\mathcal{Q}}$:
\[
\boldsymbol{\mathcal{D}}=\mathcal{\boldsymbol{\mathcal{Q}}}^{\top}\boldsymbol{\mathcal{C}}\mathcal{\boldsymbol{\mathcal{Q}}},
\]
where $\boldsymbol{\mathcal{D}}$ is a diagonal matrix (diagonal elements
are eigenvalues of $\boldsymbol{\mathcal{C}}$). Considering a change
of basis functions $\boldsymbol{\eta}^{\prime}(t)^{\top}=\boldsymbol{\mathcal{Q}}^{\top}\boldsymbol{\eta}(t)^{\top}$,
a new coefficient vector $\boldsymbol{v}^{\prime\top}=\mathcal{\boldsymbol{\mathcal{Q}}}^{\top}\boldsymbol{v}^{\top}$
has a diagonal covariance matrix:
\[
\left\langle \boldsymbol{v}^{\prime\top}\boldsymbol{v}^{\prime}\right\rangle =\left\langle \mathcal{\boldsymbol{\mathcal{Q}}}^{\top}\boldsymbol{v}^{\top}\boldsymbol{v}\mathcal{\boldsymbol{\mathcal{Q}}}\right\rangle =\mathcal{\boldsymbol{\mathcal{Q}}}^{\top}\boldsymbol{\mathcal{C}}\mathcal{\boldsymbol{\mathcal{Q}}}=\boldsymbol{\mathcal{D}},
\]
showing that elements in $\boldsymbol{v}^{\prime}$ are decorrelated.
When $\boldsymbol{v}$ obeys the multivariate Gaussian distribution,
elements in $\boldsymbol{v}^{\prime}$ are independent of each other.
Note that we cannot make arbitrary random variables independent of
each other by a change of basis functions.

\section{Optimal linear response function\label{sec:optimal_function}}

According to the Gaussian assumption of probability density of $x_{i}^{T}$
($x_{i}$ at time $t=T$), we have 
\begin{equation}
P(x_{i}^{T}|\boldsymbol{v})=\frac{1}{\sqrt{2\pi\sigma_{x_{i}}^{2}}}\exp\left(-\frac{\left(x_{i}^{T}-\sum_{j=1}^{M}v_{j}q_{ij}\right)^{2}}{2\sigma_{x_{i}}^{2}}\right).\label{eq:Pxj_v}
\end{equation}
As assumed in the main text, the probability distribution of $v_{j}$
is given by 
\begin{equation}
P(v_{j})=\frac{1}{\sqrt{2\pi\sigma_{v_{j}}^{2}}}\exp\left(-\frac{v_{j}^{2}}{2\sigma_{v_{j}}^{2}}\right).\label{eq:Pv_def}
\end{equation}

The mutual information is defined by Eq.~\eqref{eq:Ixv_def}. For
$N=M=2$ which is considered in the manuscript, with Eqs.~\eqref{eq:Pxj_v}
and \eqref{eq:Pv_def}, the mutual information $I[\boldsymbol{x}^{T};\boldsymbol{v}]$
is given by:\begin{widetext}
\begin{equation}
I[x_{1}^{T},x_{2}^{T};v_{1},v_{2}]=\frac{1}{2}\ln\left[1+\frac{q_{11}^{2}\sigma_{v_{1}}^{2}+q_{12}^{2}\sigma_{v_{2}}^{2}}{\sigma_{x_{1}}^{2}}+\frac{q_{21}^{2}\sigma_{v_{1}}^{2}+q_{22}^{2}\sigma_{v_{2}}^{2}}{\sigma_{x_{2}}^{2}}+\frac{\sigma_{v_{1}}^{2}\sigma_{v_{2}}^{2}(q_{11}q_{22}-q_{12}q_{21})^{2}}{\sigma_{x_{1}}^{2}\sigma_{x_{2}}^{2}}\right].\label{eq:Ix1x2_v1v2}
\end{equation}

\end{widetext}

We calculate optimal linear response function $h_{i}(t)$ which maximizes
the mutual information $I[\boldsymbol{x}^{T};\boldsymbol{v}]$. As
can be seen with Eq.~\eqref{eq:Ix1x2_v1v2}, the mutual information
$I[\boldsymbol{x}^{T};\boldsymbol{v}]$ is a function of $\boldsymbol{q}=[q_{ij}]$.
Instead of directly maximizing $I[\boldsymbol{x}^{T};\boldsymbol{v}]$,
we consider a more tractable function $\mathcal{M}(\boldsymbol{q})$
which satisfies the following condition: 
\[
\underset{\boldsymbol{q}}{\mathrm{argmax}}\,I[\boldsymbol{x}^{T};\boldsymbol{v}]=\underset{\boldsymbol{q}}{\mathrm{argmax}}\,\mathcal{M}(\boldsymbol{q}).
\]
Then we consider the following performance index $\mathcal{R}(\boldsymbol{q},\boldsymbol{h})$:
\begin{align}
\mathcal{R}(\boldsymbol{q},\boldsymbol{h}) & =\mathcal{M}(\boldsymbol{q})+\sum_{i,j}\lambda_{ij}\left(q_{ij}-\int_{0}^{T}h_{i}(T-t)\eta_{j}(t)dt\right)\nonumber \\
 & +\sum_{i}\Lambda_{i}\left(\sigma_{x_{i}}^{2}-2D_{i}\int_{0}^{T}h_{i}(t)^{2}dt\right),\label{eq:performance_index}
\end{align}
where $\lambda_{ij}$ and $\Lambda_{i}$ are the Lagrange multipliers.
Note that arguments of $\boldsymbol{q}$ in Eq.~\eqref{eq:performance_index}
are scalars while $\boldsymbol{h}$ are functions. Constraints corresponding
to $\lambda_{ij}$ and $\Lambda_{i}$ are derived from Eqs.~\eqref{eq:qij_def}
and \eqref{eq:conv_var}, respectively. Because $I[\boldsymbol{x}^{T};\boldsymbol{v}]$
is scale-invariant with respect to $h_{i}(t)$ and hence $\sigma_{x_{i}}$
does not affect the mutual information, we set $\sigma_{x_{i}}$ as
constant (we set $\sigma_{x_{i}}=1$ for all $i$ in the main text).
The total derivative of $\mathcal{R}(\boldsymbol{q},\boldsymbol{h})$
is written by
\begin{align}
d\mathcal{R} & =\sum_{i,j}\frac{\partial\mathcal{M}(\boldsymbol{q})}{\partial q_{ij}}dq_{ij}\nonumber \\
 & +\sum_{i,j}\lambda_{ij}\left(dq_{ij}-\int_{0}^{T}\delta h_{i}(t)\eta_{j}(T-t)dt\right)\nonumber \\
 & +\sum_{i}\Lambda_{i}\left(-4D_{i}\int_{0}^{T}h_{i}(t)\delta h_{i}(t)dt\right),\nonumber \\
 & =\sum_{i,j}\left(\frac{\partial\mathcal{M}(\boldsymbol{q})}{\partial q_{ij}}+\lambda_{ij}\right)dq_{ij}\nonumber \\
 & +\sum_{i}\int_{0}^{T}\delta h_{i}(t)\left(-\sum_{j}\lambda_{ij}\eta_{j}(T-t)-4D_{i}\Lambda_{i}h_{i}(t)\right)dt.\label{eq:dR_expr}
\end{align}
Because, $d\mathcal{R}$ should vanish at a stationary point, we obtain
the following relations: 
\begin{align}
\frac{\partial\mathcal{M}(\boldsymbol{q})}{\partial q_{ij}}+\lambda_{ij} & =0,\label{eq:dRdq}\\
-\sum_{j}\lambda_{ij}\eta_{j}(T-t)-4D_{i}\Lambda_{i}h_{i}(t) & =0.\label{eq:dRdh}
\end{align}
From Eq.~\eqref{eq:dRdh}, we obtain
\begin{equation}
h_{i}(t)=-\frac{1}{4\Lambda_{i}D_{i}}\sum_{j=1}^{M}\lambda_{ij}\eta_{j}(T-t),\label{eq:h_Opt_Appendix}
\end{equation}
which is Eq.~\eqref{eq:opt_h}. Depending on the type of decoders
(full or decorrelating), $\lambda_{ij}$ and $\Lambda_{i}$ are determined
(see below). Substituting Eq.~\eqref{eq:h_Opt_Appendix} into Eqs.~\eqref{eq:qij_def}
and \eqref{eq:conv_var}, we have 
\begin{align}
q_{ij} & =-\frac{1}{4\Lambda_{i}D_{i}}\sum_{k}\lambda_{ik}\psi_{kj},\label{eq:qij_condition}\\
\sigma_{x_{i}}^{2} & =\frac{1}{8\Lambda_{i}^{2}D_{i}}\sum_{j,k}\lambda_{ij}\lambda_{ik}\psi_{kj},\label{eq:CV_condition}
\end{align}
where $[\psi_{jj'}]$ is a correlation matrix of the basis functions
$\eta_{j}(t)$, defined by 
\[
\psi_{jj'}=\int_{0}^{T}\eta_{j}(t)\eta_{j'}(t)dt.
\]
Algebraic equations~\eqref{eq:dRdq}, \eqref{eq:qij_condition},
and \eqref{eq:CV_condition} are solved with respect to $\boldsymbol{q}$,
$\boldsymbol{\lambda}$, and $\boldsymbol{\Lambda}$ to obtain the
maximum of $I[\boldsymbol{x};\boldsymbol{v}]$.

\subsection{Full decoder}

According to Eq.~\eqref{eq:Ix1x2_v1v2}, we can use the following
function for the full decoder:
\begin{align}
\mathcal{M}(\boldsymbol{q}) & =\frac{q_{11}^{2}\sigma_{v_{1}}^{2}+q_{12}^{2}\sigma_{v_{2}}^{2}}{\sigma_{x_{1}}^{2}}+\frac{q_{21}^{2}\sigma_{v_{1}}^{2}+q_{22}^{2}\sigma_{v_{2}}^{2}}{\sigma_{x_{2}}^{2}}\nonumber \\
 & +\frac{\sigma_{v_{1}}^{2}\sigma_{v_{2}}^{2}(q_{11}q_{22}-q_{12}q_{21})^{2}}{\sigma_{x_{1}}^{2}\sigma_{x_{2}}^{2}},\label{eq:Mq_full}
\end{align}
Because it is difficult to obtain closed-form solutions for Eqs.~\eqref{eq:dRdq},
\eqref{eq:qij_condition}, and \eqref{eq:CV_condition} along with
Eq.~\eqref{eq:Mq_full}, we numerically solve the equations. 

When the noise intensity $D_{i}$ is sufficiently weak, we find the
following expression: 
\begin{align*}
I^{\mathrm{full}} & \simeq\frac{1}{2}\ln\left[\frac{\sigma_{v_{1}}^{2}\sigma_{v_{2}}^{2}(q_{11}q_{22}-q_{12}q_{21})^{2}}{\sigma_{x_{1}}^{2}\sigma_{x_{2}}^{2}}\right],\\
 & =\frac{1}{2}\ln\left[\frac{\sigma_{v_{1}}^{2}\sigma_{v_{2}}^{2}\left(\psi_{11}\psi_{22}-\psi_{12}^{2}\right)}{4D_{1}D_{2}}\right],
\end{align*}
which is Eq.~\eqref{eq:I_full_def} in the main text.

\subsection{Decorrelating decoder}

For $N=M(=2)$, which is considered in the manuscript, decorrelation
is easily implemented. The output of the $i$th decoder at time $t=T$
is denoted by $x_{i}^{T}$ and its probability density is $P(x_{i}^{T}|\boldsymbol{v})$
(Eq.~\eqref{eq:Pxj_v}). The relation can be represented by the Bayesian
network shown in Fig.~\ref{fig:eta2}(a). For this case, the output
probability density is not decorrelated, i.e., $P(\boldsymbol{x}^{T})\ne\prod_{i}P(x_{i}^{T})$
(note that since $P(\boldsymbol{x}^{T})$ is the Gaussian distribution,
decorrelation is equivalent to independence). When $P(x_{i}^{T}|\boldsymbol{v})$
disjointly depends on only one $v_{j}\in\boldsymbol{v}$ as shown
in Fig.~\ref{fig:eta2}(b), the output probability density is decorrelated.
This condition yields $q_{ij}=0$ for $i\ne j$. We can use the following
function for the decorrelating decoder: 
\begin{equation}
\mathcal{M}(\boldsymbol{q})=\frac{q_{11}^{2}\sigma_{v_{1}}^{2}}{\sigma_{x_{1}}^{2}}+\frac{q_{22}^{2}\sigma_{v_{2}}^{2}}{\sigma_{x_{2}}^{2}}+\frac{\sigma_{v_{1}}^{2}\sigma_{v_{2}}^{2}q_{11}^{2}q_{22}^{2}}{\sigma_{x_{1}}^{2}\sigma_{x_{2}}^{2}}.\label{eq:M_caseA}
\end{equation}
We obtain the mutual information as follows:\begin{widetext}
\begin{align}
I^{\mathrm{decor}} & =\frac{1}{2}\ln\left[1+\frac{\left(\psi_{11}\psi_{22}-\psi_{12}^{2}\right)\left(2D_{1}\psi_{22}\sigma_{v_{2}}^{2}+2D_{2}\psi_{11}\sigma_{v_{1}}^{2}+\left(\psi_{22}\psi_{11}-\psi_{12}^{2}\right)\sigma_{v_{1}}^{2}\sigma_{v_{2}}^{2}\right)}{4D_{1}D_{2}\psi_{11}\psi_{22}}\right].\label{eq:I_decor_exact}
\end{align}
\end{widetext}When the noise intensity $D_{i}$ is sufficiently weak,
the mutual information reduces to Eq.~\eqref{eq:I_decor_def}.

\subsection{Calculation of $I^{\mathrm{dup}}$}

Because $\eta_{1}(t)=\eta_{2}(t)=\eta(t)$ and $h_{1}(t)=h_{2}(t)=h(t)$
in $I^{\mathrm{dup}}$, $q_{ij}$ and $\sigma_{x_{i}}^{2}$ do not
depend on $i$ or $j$, where we define $q_{ij}=q$ and $\sigma_{x_{i}}^{2}=\sigma_{x}^{2}$,
respectively. Therefore, from Eq.~\eqref{eq:Ix1x2_v1v2}, the mutual
information is
\[
I[x_{1}^{T},x_{2}^{T};v_{1},v_{2}]=\frac{1}{2}\ln\left(1+\frac{4\sigma_{v}^{2}}{\sigma_{x}^{2}}q^{2}\right).
\]
Because $h(t)\propto\eta(T-t)$ from Eq.~\eqref{eq:h_Opt_Appendix},
we have $q^{2}=\psi\sigma_{x}^{2}/(2D)$ and obtain Eq.~\eqref{eq:I_dup_def_1}.

\section{Network realization of transfer function\label{sec:network_realization}}

\begin{figure}
\begin{centering}
\includegraphics[width=8cm]{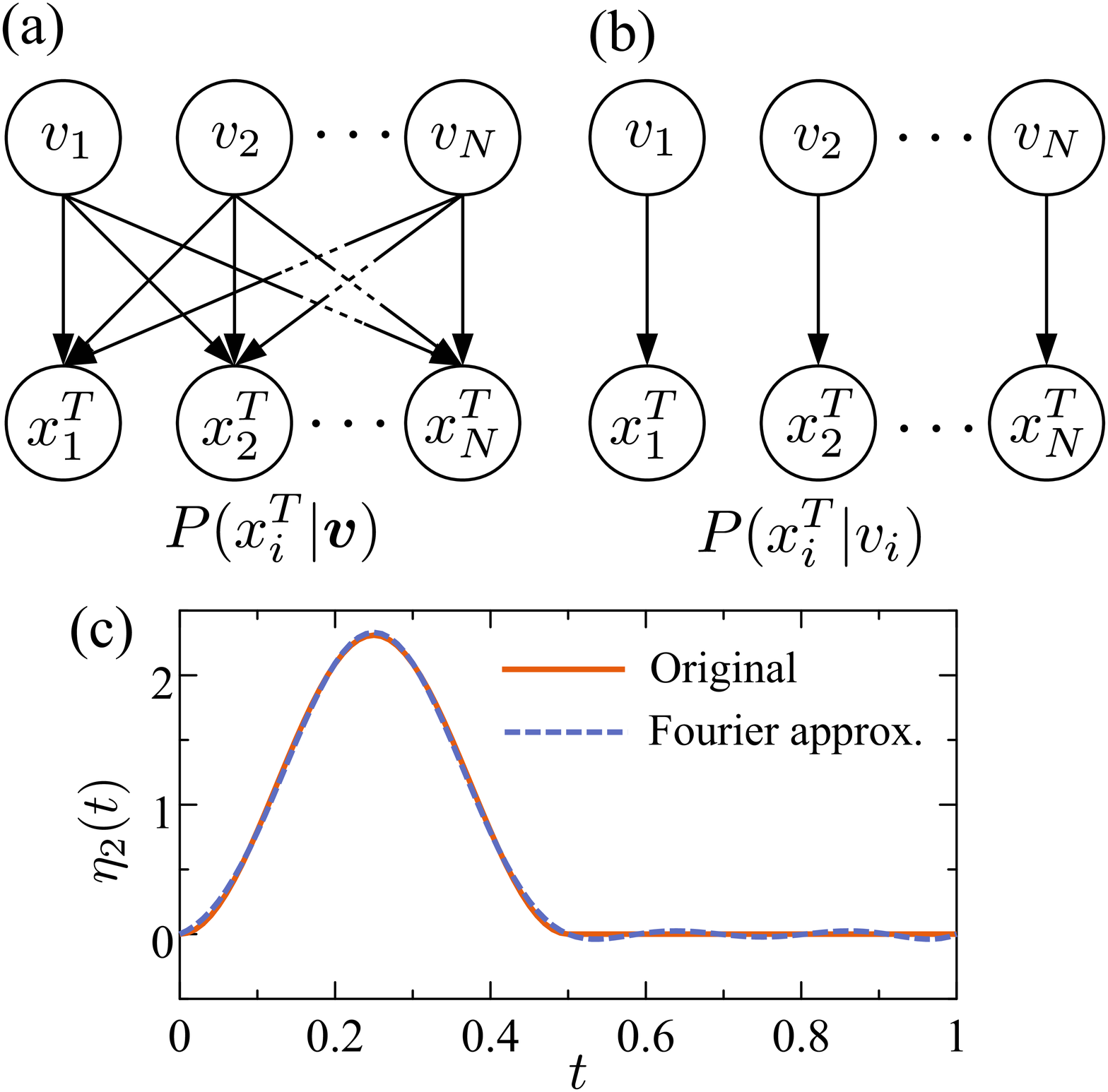} 
\par\end{centering}

\protect\caption{(a) and (b) Network representations of (a) $P(x_{i}^{T}|\boldsymbol{v})$
and (b) $P(x_{i}^{T}|v_{i})$ for $N=M$. (c) Basis function $\eta_{2}(t)$
of basis set A {[}Fig.~\ref{fig:BasisSet}(a){]} (solid line) and
its Fourier series approximation (dashed line). \label{fig:eta2}}
\end{figure}

In the main text, we explore biochemical realization of optimal linear
response functions $h_{i}(t)$. We consider a general $K$-dimensional
linear system: 
\begin{equation}
\dot{\boldsymbol{z}}(t)=\boldsymbol{A}\boldsymbol{z}(t)+\boldsymbol{b}u(t),\hspace*{1em}y(t)=\boldsymbol{c}\boldsymbol{z}(t),\label{eq:linsys_def}
\end{equation}
where $\boldsymbol{z}(t)$ is a $K$-dimensional column vector, $y(t)$
is an output scalar variable, $\boldsymbol{A}$ is a $K\times K$
matrix, $\boldsymbol{b}$ is a $K$-dimensional column vector, and
$\boldsymbol{c}$ is a $K$-dimensional row vector. Here we dropped
subscripts that identify the decoder number in order to simplify the
notation (e.g., $\boldsymbol{A}_{i}$ in the main text is simply expressed
$\boldsymbol{A}$ here) because we are describing a general theory.
It is known that the transfer function $\tilde{h}(s)$ of the linear
system of Eq.~\eqref{eq:linsys_def} is given by 
\[
\tilde{h}(s)=\boldsymbol{c}(s\boldsymbol{I}-\boldsymbol{A})^{-1}\boldsymbol{b}=\sum_{i=0}^{\infty}\frac{1}{s^{i+1}}\boldsymbol{c}\boldsymbol{A}^{i}\boldsymbol{b},
\]
where $\boldsymbol{I}$ is the identity matrix. Since the transfer
function depends only on $\boldsymbol{c}\boldsymbol{A}^{i}\boldsymbol{b}$,
the transfer function is invariant under coordinate transform $\boldsymbol{\boldsymbol{z}}^{\prime}=\mathcal{T}\boldsymbol{z}$,
where $\mathcal{T}$ is a regular matrix. According to the Faddeev
method, $(s\boldsymbol{I}-\boldsymbol{A})^{-1}$ can be calculated
by the following formula: 
\begin{align}
(s\boldsymbol{I}-\boldsymbol{A})^{-1} & =\frac{\boldsymbol{F}_{1}s^{K-1}+\cdots+\boldsymbol{F}_{K-1}s+\boldsymbol{F}_{K}}{s^{K}+f_{1}s^{K-1}+\cdots+f_{K-1}s+f_{K}},\label{eq:sI_A_inv}
\end{align}
where $\boldsymbol{F}_{i}$ and $f_{i}$ are defined as follows:
\begin{align}
\boldsymbol{F}_{1} & =\boldsymbol{I},\hspace*{1em}f_{1}=-\mathrm{tr}\boldsymbol{A},\label{eq:Faddeev1}\\
\boldsymbol{F}_{i} & =\boldsymbol{A}\boldsymbol{F}_{i-1}+f_{i-1}\boldsymbol{I},\hspace*{1em}f_{i}=-\frac{1}{i}\mathrm{tr}(\boldsymbol{A}\boldsymbol{F}_{i}).\label{eq:Faddeev2}
\end{align}
 We consider the following rational polynomial transfer function:
\begin{equation}
\tilde{h}(s)=\frac{\beta_{1}s^{K-1}+\cdots+\beta_{K-1}s+\beta_{K}}{s^{K}+\alpha_{1}s^{K-1}+\cdots+\alpha_{K-1}s+\alpha_{K}},\label{eq:tf_strictly_proper}
\end{equation}
where $\alpha_{i}$ and $\beta_{i}$ are real coefficients. One possible
realization of the transfer function of Eq.~\eqref{eq:tf_strictly_proper}
in the form of Eq.~\eqref{eq:linsys_def} is 
\begin{equation}
\boldsymbol{A}=\left[\begin{array}{ccccc}
0 & 0 & 0 & \cdots & -\alpha_{K}\\
1 & 0 & 0 & \cdots & -\alpha_{K-1}\\
0 & 1 & 0 & \cdots & -\alpha_{K-2}\\
\vdots & \vdots & \vdots & \ddots & \vdots\\
0 & 0 & 0 & 1 & -\alpha_{1}
\end{array}\right],\hspace*{1em}\boldsymbol{b}=\left[\begin{array}{c}
\beta_{K}\\
\beta_{K-1}\\
\beta_{K-2}\\
\vdots\\
\beta_{1}
\end{array}\right],\label{eq:Abc}
\end{equation}
and $\boldsymbol{c}=[\begin{array}{ccccc}
0 & 0 & \cdots & 0 & 1\end{array}]$, which is known as the observer canonical form. Because of $\boldsymbol{c}$
in Eq.~\eqref{eq:Abc}, the output is given by the last variable
$y(t)=z_{K}(t)$. 

In the main text, we consider network realization for basis set A,
whose basis functions are given in Eq.~\eqref{eq:setA1_def}. Since
the step function yields a transfer function that does not fit into
the form of Eq.~\eqref{eq:tf_strictly_proper}, we apply the Fourier
series expansion to $\eta_{2}(t)$ to obtain 
\[
\eta_{2}(t)=\frac{16\sin(2\pi t)}{3\sqrt{3}\pi}-\frac{16\sin(6\pi t)}{15\sqrt{3}\pi}-\frac{\cos(4\pi t)}{\sqrt{3}}+\frac{1}{\sqrt{3}}.
\]
In Fig.~\ref{fig:eta2}(c), we compare $\eta_{2}(t)$ of the exact
function (solid line) with the Fourier approximation (dashed line).
The Laplace transforms of $\eta_{j}(T-t)$ are given by 
\begin{align*}
\tilde{\eta}_{1}(s) & =\mathcal{L}[\eta_{1}(T-t)]=\sqrt{\frac{2}{3}}\frac{1}{s}-\sqrt{\frac{2}{3}}\frac{s}{s^{2}+4\pi^{2}},\\
\tilde{\eta}_{2}(s) & =\mathcal{L}[\eta_{2}(T-t)]=-\frac{s}{\sqrt{3}\left(s^{2}+16\pi^{2}\right)}-\frac{32}{3\sqrt{3}\left(s^{2}+4\pi^{2}\right)}\\
 & +\frac{32}{5\sqrt{3}\left(s^{2}+36\pi^{2}\right)}+\frac{1}{\sqrt{3}s},
\end{align*}
where $\mathcal{L}$ is the Laplace transform operator. From Eq.~\eqref{eq:opt_h},
the Laplace transform of optimal linear response function $h_{i}(t)$
(i.e., the transfer function) is given by Eq.~\eqref{eq:htilde_def}.
$\tilde{h}_{i}(s)$ of Eq.~\eqref{eq:htilde_def} fits into the form
of Eq.~\eqref{eq:tf_strictly_proper} since $\Lambda_{i}$ and $\lambda_{ij}$
are real values.

We next show explicit representations of $\boldsymbol{A}$ and $\boldsymbol{b}$
which are realizations of optimal linear response functions $h_{i}(t)$
($h_{1}(t)$ and $h_{2}(t)$ in Fig.~\ref{fig:Linear_MI}(a)). We
use $\boldsymbol{A}_{i}$ and $\boldsymbol{b}_{i}$ to represent $\boldsymbol{A}$
and $\boldsymbol{b}$ of the $i$th decoder:
\begin{align}
\boldsymbol{A}_{1} & =\left[\begin{array}{ccccccc}
0.0 & 0.0 & 0.0 & 0.0 & 0.0 & 0.0 & 0.0\\
10.0 & 0.0 & 0.0 & 0.0 & 0.0 & 0.0 & -22.15\\
0.0 & 10.0 & 0.0 & 0.0 & 0.0 & 0.0 & 0.0\\
0.0 & 0.0 & 10.0 & 0.0 & 0.0 & 0.0 & -76.37\\
0.0 & 0.0 & 0.0 & 10.0 & 0.0 & 0.0 & 0.0\\
0.0 & 0.0 & 0.0 & 0.0 & 10.0 & 0.0 & -55.27\\
0.0 & 0.0 & 0.0 & 0.0 & 0.0 & 10.0 & 0.0
\end{array}\right],\label{eq:realization1_A}\\
\boldsymbol{b}_{1} & =\left[\begin{array}{ccccccc}
3.78 & 1.50 & 2.32 & 1.13 & 0.35 & 0.11 & 0.0\end{array}\right]^{\top},\label{eq:realization1_b}\\
\boldsymbol{A}_{2} & =\left[\begin{array}{ccccccc}
0.0 & 0.0 & 0.0 & 0.0 & 0.0 & 0.0 & 0.0\\
10.0 & 0.0 & 0.0 & 0.0 & 0.0 & 0.0 & -22.15\\
0.0 & 10.0 & 0.0 & 0.0 & 0.0 & 0.0 & 0.0\\
0.0 & 0.0 & 10.0 & 0.0 & 0.0 & 0.0 & -76.37\\
0.0 & 0.0 & 0.0 & 10.0 & 0.0 & 0.0 & 0.0\\
0.0 & 0.0 & 0.0 & 0.0 & 10.0 & 0.0 & -55.27\\
0.0 & 0.0 & 0.0 & 0.0 & 0.0 & 10.0 & 0.0
\end{array}\right],\label{eq:realization2_A}\\
\boldsymbol{b}_{2} & =\left[\begin{array}{ccccccc}
2.25 & -7.80 & 7.93 & -5.88 & 2.05 & -0.60 & 0.0\end{array}\right]^{\top}.\label{eq:realization2_b}
\end{align}
As denoted above, these realizations are not unique, as any coordinate
transformation yields the same transfer function. Thus we applied
some scaling matrix to adjust excessively large values in Eqs.~\eqref{eq:realization1_A}--\eqref{eq:realization2_b},
which seem to be biologically infeasible. We constructed a reduced
realization for the full network of Fig.~\ref{fig:network}(a). $\boldsymbol{A}_{1}^{\prime}$
and $\boldsymbol{b}_{1}^{\prime}$, which are $\boldsymbol{A}$ and
$\boldsymbol{b}$ of the reduced network, are given by
\begin{align}
\boldsymbol{A}_{1}^{\prime} & =\left[\begin{array}{ccc}
0 & 0 & 0\\
1.0 & 0 & -39.50\\
0 & 1.0 & 0
\end{array}\right],\label{eq:Atilde_def}\\
\boldsymbol{b}_{1}^{\prime} & =\left[\begin{array}{ccc}
67.40 & 2.52 & 0\end{array}^{\top}\right].\label{eq:btilde_def}
\end{align}
Network representations of Eqs.~\eqref{eq:realization1_A}--\eqref{eq:btilde_def}
are shown in Figs.~\ref{fig:network}(a), (c), and (e) in the main
text, where positive and negative elements in $\boldsymbol{A}$ and
$\boldsymbol{b}$ are described by activation (arrow) and inhibition
(bar-headed arrow), respectively.


\begin{thebibliography}{33}
  \expandafter\ifx\csname natexlab\endcsname\relax\def\natexlab#1{#1}\fi
  \expandafter\ifx\csname bibnamefont\endcsname\relax
    \def\bibnamefont#1{#1}\fi
  \expandafter\ifx\csname bibfnamefont\endcsname\relax
    \def\bibfnamefont#1{#1}\fi
  \expandafter\ifx\csname citenamefont\endcsname\relax
    \def\citenamefont#1{#1}\fi
  \expandafter\ifx\csname url\endcsname\relax
    \def\url#1{\texttt{#1}}\fi
  \expandafter\ifx\csname urlprefix\endcsname\relax\def\urlprefix{URL }\fi
  \providecommand{\bibinfo}[2]{#2}
  \providecommand{\eprint}[2][]{\url{#2}}
  
  \bibitem[{\citenamefont{Behar and Hoffmann}(2010)}]{Behar:2010:TemporalCodes}
  \bibinfo{author}{\bibfnamefont{M.}~\bibnamefont{Behar}} \bibnamefont{and}
    \bibinfo{author}{\bibfnamefont{A.}~\bibnamefont{Hoffmann}},
    \bibinfo{journal}{Curr. Opin. Genetics Dev.} \textbf{\bibinfo{volume}{20}},
    \bibinfo{pages}{684} (\bibinfo{year}{2010}).
  
  \bibitem[{\citenamefont{Purvis and Lahav}(2013)}]{Purvis:2013:SignalingReview}
  \bibinfo{author}{\bibfnamefont{J.~E.} \bibnamefont{Purvis}} \bibnamefont{and}
    \bibinfo{author}{\bibfnamefont{G.}~\bibnamefont{Lahav}},
    \bibinfo{journal}{Cell} \textbf{\bibinfo{volume}{152}}, \bibinfo{pages}{945}
    (\bibinfo{year}{2013}).
  
  \bibitem[{\citenamefont{Kobayashi}(2010)}]{Kobayashi:2010:PRL}
  \bibinfo{author}{\bibfnamefont{T.~J.} \bibnamefont{Kobayashi}},
    \bibinfo{journal}{Phys. Rev. Lett.} \textbf{\bibinfo{volume}{104}},
    \bibinfo{pages}{228104} (\bibinfo{year}{2010}).
  
  \bibitem[{\citenamefont{Hinczewski and
    Thirumalai}(2014)}]{Hinczewski:2014:CellFilter}
  \bibinfo{author}{\bibfnamefont{M.}~\bibnamefont{Hinczewski}} \bibnamefont{and}
    \bibinfo{author}{\bibfnamefont{D.}~\bibnamefont{Thirumalai}},
    \bibinfo{journal}{Phys. Rev. X} \textbf{\bibinfo{volume}{4}},
    \bibinfo{pages}{041017} (\bibinfo{year}{2014}).
  
  \bibitem[{\citenamefont{Becker et~al.}(2015)\citenamefont{Becker, Mugler, and
    ten Wolde}}]{Becker2015:CellPrediction}
  \bibinfo{author}{\bibfnamefont{N.~B.} \bibnamefont{Becker}},
    \bibinfo{author}{\bibfnamefont{A.}~\bibnamefont{Mugler}}, \bibnamefont{and}
    \bibinfo{author}{\bibfnamefont{P.~R.} \bibnamefont{ten Wolde}},
    \bibinfo{journal}{Phys. Rev. Lett.} \textbf{\bibinfo{volume}{115}},
    \bibinfo{pages}{258103} (\bibinfo{year}{2015}).
  
  \bibitem[{\citenamefont{Kubota et~al.}(2012)\citenamefont{Kubota, Noguchi,
    Toyoshima, Ozaki, Uda, Watanabe, Ogawa, and Kuroda}}]{Kubota:2012:InsulinAKT}
  \bibinfo{author}{\bibfnamefont{H.}~\bibnamefont{Kubota}},
    \bibinfo{author}{\bibfnamefont{R.}~\bibnamefont{Noguchi}},
    \bibinfo{author}{\bibfnamefont{Y.}~\bibnamefont{Toyoshima}},
    \bibinfo{author}{\bibfnamefont{Y.-i.} \bibnamefont{Ozaki}},
    \bibinfo{author}{\bibfnamefont{S.}~\bibnamefont{Uda}},
    \bibinfo{author}{\bibfnamefont{K.}~\bibnamefont{Watanabe}},
    \bibinfo{author}{\bibfnamefont{W.}~\bibnamefont{Ogawa}}, \bibnamefont{and}
    \bibinfo{author}{\bibfnamefont{S.}~\bibnamefont{Kuroda}},
    \bibinfo{journal}{Mol. Cell} \textbf{\bibinfo{volume}{46}},
    \bibinfo{pages}{820} (\bibinfo{year}{2012}).
  
  \bibitem[{\citenamefont{Noguchi et~al.}(2013)\citenamefont{Noguchi, Kubota,
    Yugi, Toyoshima, Komori, Soga, and Kuroda}}]{Noguchi:2013:Insulin}
  \bibinfo{author}{\bibfnamefont{R.}~\bibnamefont{Noguchi}},
    \bibinfo{author}{\bibfnamefont{H.}~\bibnamefont{Kubota}},
    \bibinfo{author}{\bibfnamefont{K.}~\bibnamefont{Yugi}},
    \bibinfo{author}{\bibfnamefont{Y.}~\bibnamefont{Toyoshima}},
    \bibinfo{author}{\bibfnamefont{Y.}~\bibnamefont{Komori}},
    \bibinfo{author}{\bibfnamefont{T.}~\bibnamefont{Soga}}, \bibnamefont{and}
    \bibinfo{author}{\bibfnamefont{S.}~\bibnamefont{Kuroda}},
    \bibinfo{journal}{Mol. Syst. Biol.} \textbf{\bibinfo{volume}{9}},
    \bibinfo{pages}{664} (\bibinfo{year}{2013}).
  
  \bibitem[{\citenamefont{Sano et~al.}(2016)\citenamefont{Sano, Kawata, Ohno,
    Yugi, Kakuda, Kubota, Uda, Fujii, Kunida, Hoshino
    et~al.}}]{Sano:2016:DynPattSig}
  \bibinfo{author}{\bibfnamefont{T.}~\bibnamefont{Sano}},
    \bibinfo{author}{\bibfnamefont{K.}~\bibnamefont{Kawata}},
    \bibinfo{author}{\bibfnamefont{S.}~\bibnamefont{Ohno}},
    \bibinfo{author}{\bibfnamefont{K.}~\bibnamefont{Yugi}},
    \bibinfo{author}{\bibfnamefont{H.}~\bibnamefont{Kakuda}},
    \bibinfo{author}{\bibfnamefont{H.}~\bibnamefont{Kubota}},
    \bibinfo{author}{\bibfnamefont{S.}~\bibnamefont{Uda}},
    \bibinfo{author}{\bibfnamefont{M.}~\bibnamefont{Fujii}},
    \bibinfo{author}{\bibfnamefont{K.}~\bibnamefont{Kunida}},
    \bibinfo{author}{\bibfnamefont{D.}~\bibnamefont{Hoshino}},
    \bibnamefont{et~al.}, \bibinfo{journal}{Sci. Signal.}
    \textbf{\bibinfo{volume}{9}}, \bibinfo{pages}{ra112} (\bibinfo{year}{2016}).
  
  \bibitem[{\citenamefont{Selimkhanov et~al.}(2014)\citenamefont{Selimkhanov,
    Taylor, Yao, Pilko, Albeck, Hoffmann, Tsimring, and
    Wollman}}]{Selimkhanov:2014:DynSig}
  \bibinfo{author}{\bibfnamefont{J.}~\bibnamefont{Selimkhanov}},
    \bibinfo{author}{\bibfnamefont{B.}~\bibnamefont{Taylor}},
    \bibinfo{author}{\bibfnamefont{J.}~\bibnamefont{Yao}},
    \bibinfo{author}{\bibfnamefont{A.}~\bibnamefont{Pilko}},
    \bibinfo{author}{\bibfnamefont{J.}~\bibnamefont{Albeck}},
    \bibinfo{author}{\bibfnamefont{A.}~\bibnamefont{Hoffmann}},
    \bibinfo{author}{\bibfnamefont{L.}~\bibnamefont{Tsimring}}, \bibnamefont{and}
    \bibinfo{author}{\bibfnamefont{R.}~\bibnamefont{Wollman}},
    \bibinfo{journal}{Science} \textbf{\bibinfo{volume}{346}},
    \bibinfo{pages}{1370} (\bibinfo{year}{2014}).
  
  \bibitem[{\citenamefont{Tostevin and ten Wolde}(2009)}]{Tostevin:2009:MI}
  \bibinfo{author}{\bibfnamefont{F.}~\bibnamefont{Tostevin}} \bibnamefont{and}
    \bibinfo{author}{\bibfnamefont{P.~R.} \bibnamefont{ten Wolde}},
    \bibinfo{journal}{Phys. Rev. Lett.} \textbf{\bibinfo{volume}{102}},
    \bibinfo{pages}{218101} (\bibinfo{year}{2009}).
  
  \bibitem[{\citenamefont{Mora and Wingreen}(2010)}]{Mora:2010:MLE}
  \bibinfo{author}{\bibfnamefont{T.}~\bibnamefont{Mora}} \bibnamefont{and}
    \bibinfo{author}{\bibfnamefont{N.~S.} \bibnamefont{Wingreen}},
    \bibinfo{journal}{Phys. Rev. Lett.} \textbf{\bibinfo{volume}{104}},
    \bibinfo{pages}{248101} (\bibinfo{year}{2010}).
  
  \bibitem[{\citenamefont{Mugler et~al.}(2010)\citenamefont{Mugler, Walczak, and
    Wiggins}}]{Mugler:2010:OscSignal}
  \bibinfo{author}{\bibfnamefont{A.}~\bibnamefont{Mugler}},
    \bibinfo{author}{\bibfnamefont{A.~M.} \bibnamefont{Walczak}},
    \bibnamefont{and} \bibinfo{author}{\bibfnamefont{C.~H.}
    \bibnamefont{Wiggins}}, \bibinfo{journal}{Phys. Rev. Lett.}
    \textbf{\bibinfo{volume}{105}}, \bibinfo{pages}{058101}
    (\bibinfo{year}{2010}).
  
  \bibitem[{\citenamefont{Purvis and Lahav}(2012)}]{Purvis:2012:Insulin}
  \bibinfo{author}{\bibfnamefont{J.~E.} \bibnamefont{Purvis}} \bibnamefont{and}
    \bibinfo{author}{\bibfnamefont{G.}~\bibnamefont{Lahav}},
    \bibinfo{journal}{Mol. Cell} \textbf{\bibinfo{volume}{46}},
    \bibinfo{pages}{715} (\bibinfo{year}{2012}).
  
  \bibitem[{\citenamefont{Hansen and O'Shea}(2013)}]{Hansen:2013:DynDec}
  \bibinfo{author}{\bibfnamefont{A.~S.} \bibnamefont{Hansen}} \bibnamefont{and}
    \bibinfo{author}{\bibfnamefont{E.~K.} \bibnamefont{O'Shea}},
    \bibinfo{journal}{Mol. Syst. Biol.} \textbf{\bibinfo{volume}{9}},
    \bibinfo{pages}{704} (\bibinfo{year}{2013}).
  
  \bibitem[{\citenamefont{Behar et~al.}(2013)\citenamefont{Behar, Barken, Werner,
    and Hoffmann}}]{Behar:2013:Dynamics}
  \bibinfo{author}{\bibfnamefont{M.}~\bibnamefont{Behar}},
    \bibinfo{author}{\bibfnamefont{D.}~\bibnamefont{Barken}},
    \bibinfo{author}{\bibfnamefont{S.~L.} \bibnamefont{Werner}},
    \bibnamefont{and} \bibinfo{author}{\bibfnamefont{A.}~\bibnamefont{Hoffmann}},
    \bibinfo{journal}{Cell} \textbf{\bibinfo{volume}{155}}, \bibinfo{pages}{448}
    (\bibinfo{year}{2013}).
  
  \bibitem[{\citenamefont{Mc~Mahon et~al.}(2015)\citenamefont{Mc~Mahon, Lenive,
    Filippi, and Stumpf}}]{McMahon:2015:MI}
  \bibinfo{author}{\bibfnamefont{S.~S.} \bibnamefont{Mc~Mahon}},
    \bibinfo{author}{\bibfnamefont{O.}~\bibnamefont{Lenive}},
    \bibinfo{author}{\bibfnamefont{S.}~\bibnamefont{Filippi}}, \bibnamefont{and}
    \bibinfo{author}{\bibfnamefont{M.~P.~H.} \bibnamefont{Stumpf}},
    \bibinfo{journal}{J. R. Soc. Interface} \textbf{\bibinfo{volume}{12}},
    \bibinfo{pages}{20150597} (\bibinfo{year}{2015}).
  
  \bibitem[{\citenamefont{Makadia et~al.}(2015)\citenamefont{Makadia, Schwaber,
    and Vadigepalli}}]{Makadia:2015:DynamicSignal}
  \bibinfo{author}{\bibfnamefont{H.~K.} \bibnamefont{Makadia}},
    \bibinfo{author}{\bibfnamefont{J.~S.} \bibnamefont{Schwaber}},
    \bibnamefont{and}
    \bibinfo{author}{\bibfnamefont{R.}~\bibnamefont{Vadigepalli}},
    \bibinfo{journal}{PLoS Comput. Biol.} \textbf{\bibinfo{volume}{11}},
    \bibinfo{pages}{e1004563} (\bibinfo{year}{2015}).
  
  \bibitem[{\citenamefont{de~Ronde et~al.}(2011)\citenamefont{de~Ronde, Tostevin,
    and ten Wolde}}]{Ronde:2011:Multiplex}
  \bibinfo{author}{\bibfnamefont{W.}~\bibnamefont{de~Ronde}},
    \bibinfo{author}{\bibfnamefont{F.}~\bibnamefont{Tostevin}}, \bibnamefont{and}
    \bibinfo{author}{\bibfnamefont{P.~R.} \bibnamefont{ten Wolde}},
    \bibinfo{journal}{Phys. Rev. Lett.} \textbf{\bibinfo{volume}{107}},
    \bibinfo{pages}{048101} (\bibinfo{year}{2011}).
  
  \bibitem[{\citenamefont{de~Ronde and ten Wolde}(2014)}]{Ronde:2014:Multiplex}
  \bibinfo{author}{\bibfnamefont{W.}~\bibnamefont{de~Ronde}} \bibnamefont{and}
    \bibinfo{author}{\bibfnamefont{P.~R.} \bibnamefont{ten Wolde}},
    \bibinfo{journal}{Phys. Biol.} \textbf{\bibinfo{volume}{11}},
    \bibinfo{pages}{026004} (\bibinfo{year}{2014}).
  
  \bibitem[{\citenamefont{Gallager}(1968)}]{Gallager:1968:InfoTheory}
  \bibinfo{author}{\bibfnamefont{R.~G.} \bibnamefont{Gallager}},
    \emph{\bibinfo{title}{Information theory and reliable communication}},
    vol.~\bibinfo{volume}{2} (\bibinfo{publisher}{Springer},
    \bibinfo{year}{1968}).
  
  \bibitem[{\citenamefont{Wang et~al.}(2007)\citenamefont{Wang, Rappel, Kerr, and
    Levine}}]{Wang:2007:QuantifyNoise}
  \bibinfo{author}{\bibfnamefont{K.}~\bibnamefont{Wang}},
    \bibinfo{author}{\bibfnamefont{W.-J.} \bibnamefont{Rappel}},
    \bibinfo{author}{\bibfnamefont{R.}~\bibnamefont{Kerr}}, \bibnamefont{and}
    \bibinfo{author}{\bibfnamefont{H.}~\bibnamefont{Levine}},
    \bibinfo{journal}{Phys. Rev. E} \textbf{\bibinfo{volume}{75}},
    \bibinfo{pages}{061905} (\bibinfo{year}{2007}).
  
  \bibitem[{\citenamefont{Govern and ten
    Wolde}(2012)}]{Govern:2012:LinearNetwork}
  \bibinfo{author}{\bibfnamefont{C.~C.} \bibnamefont{Govern}} \bibnamefont{and}
    \bibinfo{author}{\bibfnamefont{P.~R.} \bibnamefont{ten Wolde}},
    \bibinfo{journal}{Phys. Rev. Lett.} \textbf{\bibinfo{volume}{109}},
    \bibinfo{pages}{218103} (\bibinfo{year}{2012}).
  
  \bibitem[{\citenamefont{Marhl et~al.}(2006)\citenamefont{Marhl, Perc, and
    Schuster}}]{Marhl:2006:CaDecoder}
  \bibinfo{author}{\bibfnamefont{M.}~\bibnamefont{Marhl}},
    \bibinfo{author}{\bibfnamefont{M.}~\bibnamefont{Perc}}, \bibnamefont{and}
    \bibinfo{author}{\bibfnamefont{S.}~\bibnamefont{Schuster}},
    \bibinfo{journal}{Biophys. Chem.} \textbf{\bibinfo{volume}{120}},
    \bibinfo{pages}{161} (\bibinfo{year}{2006}).
  
  \bibitem[{\citenamefont{Hasegawa and
    Arita}(2014{\natexlab{a}})}]{Hasegawa:2013:OptimalPRC}
  \bibinfo{author}{\bibfnamefont{Y.}~\bibnamefont{Hasegawa}} \bibnamefont{and}
    \bibinfo{author}{\bibfnamefont{M.}~\bibnamefont{Arita}}, \bibinfo{journal}{J.
    R. Soc. Interface} \textbf{\bibinfo{volume}{11}}, \bibinfo{pages}{20131018}
    (\bibinfo{year}{2014}{\natexlab{a}}).
  
  \bibitem[{\citenamefont{Hasegawa and
    Arita}(2014{\natexlab{b}})}]{Hasegawa:2014:PRL}
  \bibinfo{author}{\bibfnamefont{Y.}~\bibnamefont{Hasegawa}} \bibnamefont{and}
    \bibinfo{author}{\bibfnamefont{M.}~\bibnamefont{Arita}},
    \bibinfo{journal}{Phys. Rev. Lett.} \textbf{\bibinfo{volume}{113}},
    \bibinfo{pages}{108101} (\bibinfo{year}{2014}{\natexlab{b}}).
  
  \bibitem[{\citenamefont{Hasegawa}(2016)}]{Hasegawa:2016:DynSignal}
  \bibinfo{author}{\bibfnamefont{Y.}~\bibnamefont{Hasegawa}},
    \bibinfo{journal}{New J. Phys.} \textbf{\bibinfo{volume}{18}},
    \bibinfo{pages}{113031} (\bibinfo{year}{2016}).
  
  \bibitem[{\citenamefont{Williams and Lawrence}(2007)}]{Williams:2007:LSS}
  \bibinfo{author}{\bibfnamefont{R.~L.} \bibnamefont{Williams}} \bibnamefont{and}
    \bibinfo{author}{\bibfnamefont{D.~A.} \bibnamefont{Lawrence}},
    \emph{\bibinfo{title}{Linear state-space control systems}}
    (\bibinfo{publisher}{John Wiley \& Sons}, \bibinfo{year}{2007}).
  
  \bibitem[{\citenamefont{Mangan and Alon}(2003)}]{Mangan:2003:FFL}
  \bibinfo{author}{\bibfnamefont{S.}~\bibnamefont{Mangan}} \bibnamefont{and}
    \bibinfo{author}{\bibfnamefont{U.}~\bibnamefont{Alon}},
    \bibinfo{journal}{Proc. Natl. Acad. Sci. U.S.A.}
    \textbf{\bibinfo{volume}{100}}, \bibinfo{pages}{11980}
    (\bibinfo{year}{2003}).
  
  \bibitem[{\citenamefont{Alon}(2007)}]{Alon:2007:SystBiolBook}
  \bibinfo{author}{\bibfnamefont{U.}~\bibnamefont{Alon}},
    \emph{\bibinfo{title}{An Introduction to Systems Biology}}
    (\bibinfo{publisher}{CRC Press}, \bibinfo{year}{2007}).
  
  \bibitem[{\citenamefont{Parrondo et~al.}(2015)\citenamefont{Parrondo, Horowitz,
    and Sagawa}}]{Parrondo:2015:InfoThermo}
  \bibinfo{author}{\bibfnamefont{J.~M.~R.} \bibnamefont{Parrondo}},
    \bibinfo{author}{\bibfnamefont{J.~M.} \bibnamefont{Horowitz}},
    \bibnamefont{and} \bibinfo{author}{\bibfnamefont{T.}~\bibnamefont{Sagawa}},
    \bibinfo{journal}{Nat. Phys.} \textbf{\bibinfo{volume}{11}},
    \bibinfo{pages}{131} (\bibinfo{year}{2015}).
  
  \bibitem[{\citenamefont{Sartori et~al.}(2014)\citenamefont{Sartori, Granger,
    Lee, and Horowitz}}]{Sartori:2014:Adaptation}
  \bibinfo{author}{\bibfnamefont{P.}~\bibnamefont{Sartori}},
    \bibinfo{author}{\bibfnamefont{L.}~\bibnamefont{Granger}},
    \bibinfo{author}{\bibfnamefont{C.~F.} \bibnamefont{Lee}}, \bibnamefont{and}
    \bibinfo{author}{\bibfnamefont{J.~M.} \bibnamefont{Horowitz}},
    \bibinfo{journal}{PLoS Comput. Biol.} \textbf{\bibinfo{volume}{10}},
    \bibinfo{pages}{e1003974} (\bibinfo{year}{2014}).
  
  \bibitem[{\citenamefont{Barato et~al.}(2014)\citenamefont{Barato, Hartich, and
    Seifert}}]{Barato:2014:CellInfo}
  \bibinfo{author}{\bibfnamefont{A.~C.} \bibnamefont{Barato}},
    \bibinfo{author}{\bibfnamefont{D.}~\bibnamefont{Hartich}}, \bibnamefont{and}
    \bibinfo{author}{\bibfnamefont{U.}~\bibnamefont{Seifert}},
    \bibinfo{journal}{New J. Phys.} \textbf{\bibinfo{volume}{16}},
    \bibinfo{pages}{103024} (\bibinfo{year}{2014}).
  
  \bibitem[{\citenamefont{Ouldridge et~al.}(2017)\citenamefont{Ouldridge, Govern,
    and ten Wolde}}]{Ouldridge:2017:ThermoCost}
  \bibinfo{author}{\bibfnamefont{T.~E.} \bibnamefont{Ouldridge}},
    \bibinfo{author}{\bibfnamefont{C.~C.} \bibnamefont{Govern}},
    \bibnamefont{and} \bibinfo{author}{\bibfnamefont{P.~R.} \bibnamefont{ten
    Wolde}}, \bibinfo{journal}{Phys. Rev. X} \textbf{\bibinfo{volume}{7}},
    \bibinfo{pages}{021004} (\bibinfo{year}{2017}).
  
  \end{thebibliography}
\end{document}